\documentclass[
aps,
prl,
showpacs
amsmath,
amssymb,
onecolumn,
floatfix,
letterpaper,
lengthcheck,
superscriptaddress
]{revtex4-2}

\usepackage{graphicx}
\usepackage{floatflt,epsfig} 
\usepackage{mathtools}
\usepackage{color}
\usepackage{braket}
\usepackage{hyperref}
\usepackage[normalem]{ulem}
\usepackage[english]{babel}
\usepackage{blindtext}
\usepackage{lipsum}  
\usepackage{amsthm}
\usepackage{tabularx}
\usepackage{bm}
\usepackage{dsfont}
\usepackage{bbold}
\usepackage{ulem}
\usepackage{mwe,tikz}\usepackage[percent]{overpic}

\begin{document}

\title{A Shortcut to Self-Consistent Light-Matter Interaction and Realistic Spectra from First-Principles}

  \author{Christian Sch\"afer}
  \email[Electronic address:\;]{christian.schaefer.physics@gmail.com}
  \affiliation{Department of Microtechnology and Nanoscience, MC2, Chalmers University of Technology, 412 96 G\"oteborg, Sweden}
  \affiliation{Department of Physics, Chalmers University of Technology, 412 96 G\"oteborg, Sweden}  
  \author{G\"oran Johansson}
  \affiliation{Department of Microtechnology and Nanoscience, MC2, Chalmers University of Technology, 412 96 G\"oteborg, Sweden}
  \affiliation{Department of Physics, Chalmers University of Technology, 412 96 G\"oteborg, Sweden}  
  
\date{\today}

\begin{abstract}
We introduce a simple approach how an electromagnetic environment can be efficiently embedded into state-of-the-art electronic structure methods, taking the form of radiation-reaction forces. We demonstrate that this self-consistently provides access to radiative emission, natural linewidth, Lamb shifts, strong-coupling, electromagnetically induced transparency, Purcell-enhanced and superradiant emission. As an example, we illustrate its seamless integration into time-dependent density-functional theory with virtually no additional cost, presenting a convenient shortcut to light-matter interactions.
\end{abstract}

\date{\today}

\maketitle


The theoretical description of light interacting with realistic materials becomes increasingly interdisciplinary due to the recent developments in non-equilibrium phase-transitions \cite{li2018theory,li2019terahertz,shin2021simulating}, Floquet engineering \cite{sentef2015,huebener2017,hubener2021engineering}, high-harmonic generation \cite{tancogne2017impact,floss2018ab}, strong light-matter coupling and induced modifications of energy transfer \cite{orgiu2015,schachenmayer2015,zhong2016,du2018theory,schafer2019modification,saez2018organic,groenhof2019tracking,kockum2019ultrastrong,wang2017coherent,rossi2019strong}, polaritonic chemistry \cite{hutchison2012,thomas2016,herrera2016cavity,campos2019resonant,schafer2021shining,li2021cavity,li2021collective}, plasmonic strong coupling \cite{kuhn2006enhancement,halas2011plasmons,marinica2015active,chikkaraddy2016,fregoni2018manipulating,munkhbat2018suppression,franke2019quantization,neuman2020nanomagnonic}, \textit{ab initio} QED \cite{flick2017b,schafer2018insights,ruggenthaler2018quantum,haugland2020coupled,haugland2020intermolecular,schafer2021making} and quantum-electrodynamical density-functional theory (QEDFT) \cite{ruggenthaler2014,tokatly2013,PhysRevLett.121.113002,schafer2021making}. It is apparent that all light-matter interactions arise from the interplay of matter with a electromagnetic environment. The involved material is routinely described using \textit{ab initio} electronic structure theory. A notable representative is time-dependent density-functional theory (TDDFT) \cite{gross1984,van1999mapping} due to its good accuracy and comparably low computational cost.  While it is common to consider the electromagnetic field as input to the system, i.e., a laser driving the system, the interaction should be treated self-consistently within the \textit{ab initio} calculation according to Maxwell's equations. 
This avoids negative side effects such as artificial heating but as we will show here, also naturally introduces a plethora of phenomena such as natural linewidth, Lamb shift, Purcell enhancement \cite{purcell1995spontaneous}, superradiance \cite{gross1982superradiance,goban2015}, strong light-matter coupling and electromagnetically induced transparency \cite{fleischhauer2005electromagnetically}. These are common objectives for quantum optics \cite{rivas2012open} and open quantum-system dynamics but are rarely even considered in state-of-the-art \textit{ab initio} calculations. 
Existing open-system extensions of density-functional theory \cite{yuen2010time} are thus far limited in their applicability due to physically less motivated \cite{albrecht1975new} or much more involved constructions \cite{zheng2007time,tokatly2013,jestadt2018real}. 
The TDDFT codes Salmon \cite{noda2019salmon} and Octopus \cite{jestadt2018real,tancogne2019octopus} allow the self-consistent propagation of Maxwell and Kohn-Sham equations but the complexity of the implementation and its computational cost limit their widespread use.
Only recent developments along the lines of Mixed-Quantum classical techniques \cite{hoffmann2019benchmarking}, dissipative equations of motion \cite{bustamante2021dissipative} and Casida QEDFT \cite{flick2018light} started to address natural lifetimes from first principles.
We demonstrate in the following a computationally and conceptually simpler way to extend \textit{ab initio} electronic structure approaches with electromagnetic emission, taking the form of radiation-reaction forces.
By embedding Maxwell's equation of motion into the ordinary TDDFT routine via a simple local potential, we obtain access to open quantum-system dynamics from first principles with virtually no additional effort. 
Its precise strength is that it allows the description of the aforementioned physical effects without the need to change the existing libraries.

\textit{Embedding the electromagnetic environment - }
The non-relativistic dynamic of matter subject to a classical electromagnetic environment is governed by the minimally coupled Coulomb Hamiltonian
$ \hat{H} = \sum_i \frac{1}{2m_i}\left( -i\hbar\nabla_i - q_i \textbf{A}(\textbf{r}_i t)/c \right)^2 + \hat{H}_\parallel + \varepsilon_0/2\int d\textbf{r}[ \textbf{E}_{\perp}(\textbf{r}t)^2 + c^2\textbf{B}(\textbf{r}t)^2] $ with fixed Coulomb gauge $\nabla \cdot \textbf{A}=0$. The electromagnetic fields obey Maxwell's equation of motion and couple self-consistently with the Schr\"odinger equation. The internal structure of the latter, consisting out of electrons and nuclei, is determined by the longitudinal Coulombic interactions  $\hat{H}_\parallel=1/(8\pi\varepsilon_0)\sum^{N_e+N_n}_{i, j} q_iq_j/\vert \textbf{r}_i-\textbf{r}_j\vert$, here given in free-space. The transverse electric $\textbf{E}_\perp(\textbf{r}t)=-1/c\partial_t \textbf{A}(\textbf{r}t)$ and magnetic  fields $\textbf{B}(\textbf{r}t)=1/c\nabla \times \textbf{A}(\textbf{r}t)$ are in contrast not bound to the material but can propagate into free-space. 
Even with the restriction to classical electromagnetic fields we will obtain access to the various highlighted quantum optical effects, while for instance photon blockade \cite{birnbaum2005photon} will demand quantized fields.
If the wavelength of those propagating fields is substantially larger than the matter system, the spatial dependence of the transverse fields is commonly neglected (dipole approximation). Even in this strongly simplified limit, the remaining task of solving the Schr\"odinger equation is extraordinarily challenging. We will fix the nuclei in the following for brevity but would like to emphasize that their contribution can be easily reinstated by using the total (electronic plus nuclear) dipole. Nuclear motion itself is evidently often a relevant feature to recover experimental spectra.

After performing the Power-Zienau-Wooley transformation \cite{power1959coulomb,schafer2019relevance} (see SI), the classical transverse light-matter coupling acting on the electronic structure takes the bilinear form $\hat{V}(t) =-\hat{\textbf{R}}\cdot\textbf{E}_\perp(t)$ with the electronic dipole $\hat{\textbf{R}}=-e\sum_i^{N_e} \textbf{r}_i$. The electric field can be separated into driving fields $\textbf{E}_{\text{drive},\perp}$ of external origin and radiated fields $\textbf{E}_{r,\perp}$ that are generated by the material itself. 
Naturally, driving the system externally will lead to absorption, excitation and heat.
In principle, we could solve Schr\"odinger and Maxwell equation now side by side in order to obtain access to the radiated fields and with it to the self-consistent evolution of light and matter. Given that electrons and fields move at different speeds and scales, this introduces however an unnecessary complexity that can be circumvented by the following shortcut.

Each current induces an electromagnetic field for which its precise spatial and polarization structure depends on the electromagnetic environment -- oscillating charges emit light. The generated field can be expressed with the help of the dyadic Green's tensor $\textbf{G}$ 
\begin{align*}
\textbf{E}_r(\textbf{r},\omega) = i\mu_0\omega\int_V dr' \textbf{G}(\textbf{r},\textbf{r}',\omega) \cdot (-e\textbf{j}(\textbf{r}',\omega))
\end{align*}
which is the formal solution of Helmholtz's equation \cite{buhmann2013dispersionI}
\begin{align*}
\big[\nabla \times \frac{1}{\mu_r(\textbf{r}\omega)}\nabla \times - \omega^2 \mu_0\varepsilon_0\varepsilon_r(\textbf{r}\omega)\big] \textbf{G}(\textbf{r},\textbf{r}',\omega) = \boldsymbol\delta(\textbf{r},\textbf{r}')
\end{align*}
and characterizes the electromagnetic environment. The latter comprises foremost the boundary conditions of the field but can furthermore account for
 linear media $\varepsilon_r,~\mu_r$. This provides the flexibility to describe parts of the entire matter-system in a simplified fashion  while we focus our computational effort on the electronic structure considered via the paramagnetic current density $\textbf{j}$. Such a separation becomes particularly interesting in multi-component systems that extend over various length-scales, e.g. molecule (described microscopically via $\textbf{j}$) and solvent (described macroscopically via $\varepsilon_r$). We will highlight this in more detail in a subsequent publication. As long as we treat the full system explicitly via the microscopic currents $\textbf{j}$ (as exemplified in the following), no limitation to the field strength applies. The introduction of linear media on the other hand limits the evaluation to the linear response regime. The via $\textbf{G}$ embedded environment is obtained either analytically or numerically, a standard task for Maxwell solvers, and provides the realistic mode-structure to which the electronic structure calculations couple. 
The energy associated with the radiated field should be taken correctly from the electronic system. Hence, the system of oscillating charges should feel a recoil force that accounts for any emitted energy -- the radiation reaction, ensuring Newtons third law. Consistent with the above bilinear coupling, the (dipolar) radiation-reaction potential $ \hat{V}_{rr}(t)=-\hat{\textbf{R}}\cdot\textbf{E}_{r,\perp}(t) =-\hat{\textbf{R}}\cdot \big[\mathcal{F}_t^{-1}(i\mu_0 \omega \textbf{G}_\perp (\omega)) \ast \int dr (-e \textbf{j}(\textbf{r}t))\big]  $, accounts now for the self-consistent interaction and the loss of energy due to photonic emission $\Delta E_{rr}(t)= \int_{t_0}^{t} dt'  \dot{\textbf{R}}(t') \cdot \textbf{E}_{r,\perp}(t') $. 
We notice that $ \hat{V}_{rr}(t) \propto  \hat{\textbf{R}} \cdot \textbf{G}_\perp \cdot \textbf{j} $, i.e., the transverse projection of the paramagnetic current induces a (memory dependent) recoil acting back on the electronic system.
In the case of free-space, the radiation-reaction potential takes the form $\hat{V}_{rr}(t) = \frac{-1}{6\pi\varepsilon_0c^3} \partial_t^3 \textbf{R}(t) \cdot \hat{\textbf{R}}$, which is consistent with the classical Abraham-Lorentz model \cite{wheeler1945interaction,jackson1999classical} (more details in the SI). Calculating $\textbf{G}$ up-front, the self-consistency is embedded solely via the current which avoids the need to treat electronic structure and electromagnetic fields at the same time. The simplicity of the radiation-reaction potential represents the major strength of this approach, providing a swift and intuitive usage in a variety of electronic structure libraries.

Clearly, $\textbf{G}$ and any emerging parameters depend on the electromagnetic environment of our choice.
Let us illustrate this conception in more detail for a strongly idealized one-dimensional waveguide ($\varepsilon_r=\mu_r=1$, with periodic boundaries in emission direction). 
The solution to Helmholtz's equation is then $\textbf{G}_\perp (x,x',\omega) = \sum_{\textbf{k}} \frac{S(x) S(x') }{k^2-(\omega/c)^2} \boldsymbol\epsilon_c \boldsymbol\epsilon_c^T$ with the eigemodes $S(\textbf{r}) = \sqrt{1/V}(\cos(k x)+\sin(k x)),~k=2\pi n_x/L_x,~n_x\in \mathbb{Z}$ and $\boldsymbol\epsilon_c \perp \textbf{k}$. In combination with the inverse Fourier transformation and performing the long-wavelength approximation ($x=x'=0$, details in the SI), the radiation-reaction potential takes the form 
$\hat{V}_{rr}(t) = -\hat{\textbf{R}}\cdot \boldsymbol\epsilon_c  \sum_{\textbf{k}} \frac{1}{V\varepsilon_0}\int_{-\infty}^{t} dt'\cos(ck(t-t')) 
\int_V dr'  \boldsymbol\epsilon_c\cdot (-e\textbf{j}(\textbf{r}'t'))$. 
When the number of photonic modes is increased ($L_x\rightarrow \infty$), the dense mode-structure will start to represent a photonic bath. 
Performing the explicit integration and employing the continuity equation for the integrated current $ \int dr (-e\textbf{j}(\textbf{r}t)) = \partial_t\textbf{R}(t) $, we obtain 
the radiation-reaction potential accounting for the recoil forces of emitting into the simplified waveguide
$\hat{V}^{1D}_{rr}(t) = \frac{4\pi\hbar\alpha}{e^2}A^{-1}~ \boldsymbol\epsilon_c \cdot \dot{\textbf{R}}(t) ~\boldsymbol\epsilon_c \cdot \hat{\textbf{R}}$.
The coupling between the photonic continuum (representing the perfect wide-band limit) and the electronic system is provided by the fine-structure constant $\alpha$ divided by the cross-sectional area of the waveguide $A = V/L_x$. 
The integrated emitted energy takes the form $\Delta E_{rr}(t)=\frac{4\pi\hbar\alpha}{e^2}A^{-1} \int_{t_0}^{t} dt' \vert \boldsymbol\epsilon_c \cdot\dot{\textbf{R}}(t') \vert^2 $.

While {\aa}ngstr{\"o}m thickness waveguides are possible \cite{zhang2019guiding}, the majority of interesting systems (such as typical Fabry-Perot-type cavities) will feature comparably large quantization volumes, leading to a small influence of the radiation-reaction on the dynamics. Plasmonic systems represent here an exception due to their large currents and high mode-density at nanometer scales \cite{ozbay2006plasmonics,koenderink2010use,nielsen2017giant} which substantially enhances the emission of nearby molecular systems (Purcell-enhancement) as discussed later.
An extended discussion and first generalizations to free-space and emission near mirrors can be found in the SI, we will remain here with our illustrative example.

\begin{figure}
\includegraphics[width=1.0\columnwidth]{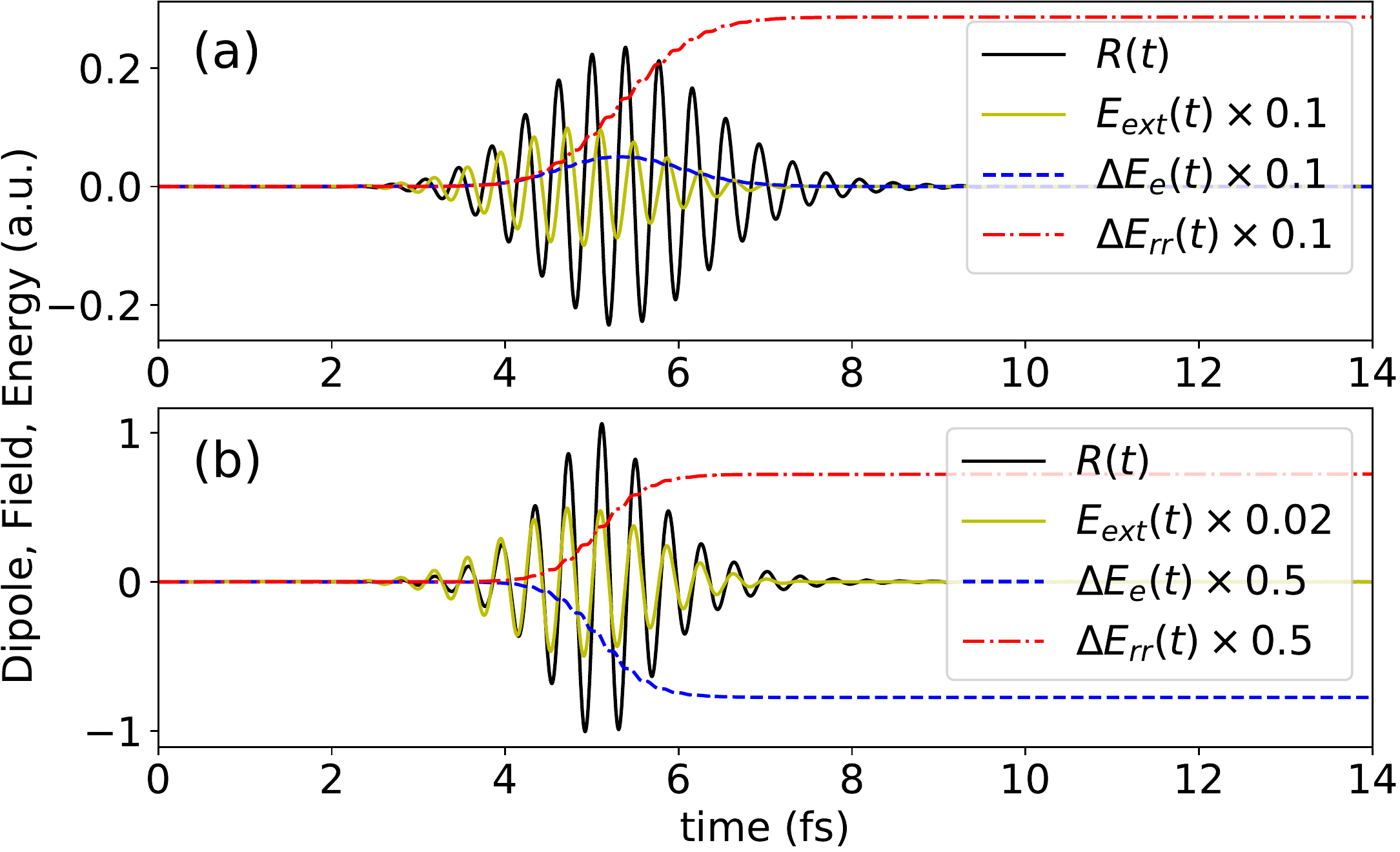}
\caption{Dipole $R(t)$ (black solid), relative electronic energy $\Delta E_e$ (blue dashed) and accumulated emitted radiation-reaction energy $\Delta E_{rr}$ (red dashed-dotted) for one-dimensional soft-coulomb hydrogen $v_{ext}(x)=e^2/(4\pi\varepsilon_0 \sqrt{x^2+1})$ emitting into a waveguide assuming a strongly Purcell-enhanced emission with $A^{-1}=1/a_0^2$ for illustrative reasons. Notice that Purcell-enhancement factors of order $\Gamma/\Gamma_0=\mathcal{O}(10^5)$ are not unusual. (a) An external laser-pulse $E_{ext}(t)$ (yellow, solid, amplified by a factor 10) drives the system out of its ground-state. Energy deposited in the electronic system  by the pulse is quickly radiated into the photonic bath. (b) Stimulated emission from the first excited electronic state under external driving. Details in SI.}
\label{fig:hydrogen1d_energy_dipole}
\end{figure}

\textit{TDDFT as example and self-consistent emission - }
Time-dependent density-functional theory evolves around the concept that effective single-particle Kohn-Sham equations exist that are able to uniquely mimic all physical observables which are inherited by the original Coulomb Hamiltonian \cite{gross1984,van1999mapping,engel2013density}.
Our goal is now to provide a local potential that is as simple as possible to describe light-matter coupling and radiative emission consistent in the ordinary time-dependent Kohn-Sham equations
$i\hbar\partial_t \phi_i(\textbf{r}t) = [-\frac{\hbar^2}{2m_e}\nabla^2 + v_{s}(\textbf{r}t)] \phi_i(\textbf{r}t)$.
The local Kohn-Sham potential 
$v_{s} = v_{ext} + v_{Hxc} + v_{rr}$ 
consists of the external potential $v_{ext}$ (usually the nuclear binding potential), the electronic Hartree-exchange-correlation potential $v_{Hxc}$ mimicking electronic many-body interactions, and the radiation-reaction potential $v_{rr}$ accounts for the coupling to light. For each single-particle equation, the dipole-operator separates into single-coordinate contributions such that the radiation-reaction potential takes the trivial form
\begin{align}
\label{eq:rrpot1d}
v^{1D}_{rr}(\textbf{r}t) &= \frac{4\pi\hbar\alpha}{e^2}A^{-1}~ \boldsymbol\epsilon_c \cdot \dot{\textbf{R}}(t) ~\boldsymbol\epsilon_c \cdot (-e\textbf{r})~.
\end{align}
The potential can be added with virtually no effort to any TDDFT routine and accounts now consistently for excitation and radiative de-excitation according to Maxwell's equation of motion (see the SI for an alternative derivation from QEDFT).
Figure~\ref{fig:hydrogen1d_energy_dipole} exemplifies the quick radiative decay of hydrogen driven by an external pulse (a) and the process of stimulated emission (b). Also strong-field effects such as high-harmonic generation can be described with the radiation-reaction potential (see SI).


External perturbations $\delta v_{appl}$ induce a response of the system which results in electronic motion and thus emission according to the radiation-reaction forces. In linear order, the response of the electronic density is given by
$\delta \rho (\textbf{r}t) = \int dr' dt' \chi_s(\textbf{r},\textbf{r}',t-t')  [\delta v_{appl}(\textbf{r}'t') 
+ \int dr'' dt'' f(\textbf{r}',\textbf{r}'',t'-t'') \delta \rho (\textbf{r}''t'')
]~.$
Here $f(\textbf{r}',\textbf{r}'',t'-t'') = \delta v_{s}(\textbf{r}'t')/\delta\rho(\textbf{r}''t'')$ is the kernel which accounts for the linear response of the Kohn-Sham potential (which depends self-consistently on the density) when the electronic density is perturbed. The contribution of the radiation-reaction forces
$f^{1D}_{rr} (\textbf{r}',\textbf{r}'',\omega) = i 4\pi\hbar\alpha A^{-1} \omega \boldsymbol\epsilon_c \cdot\textbf{r}' \boldsymbol\epsilon_c \cdot\textbf{r}''$
is explicitly memory- and thus frequency-dependent which allows it to provide a broadened resonance in contrast to the widely used frequency-independent adiabatic kernels that demand \textit{ad hoc} broadening by hand. 
Assuming a single occupied (g) and a single unoccupied (e) Kohn-Sham orbital with bare excitation energy $\Omega_{eg} = \hbar\omega_{eg}= \varepsilon_e-\varepsilon_g$, the linear-response Casida equation can be solved analytically. We obtain the excitation poles
$
\Omega_{n} = \pm \Omega_{eg} [ \sqrt{1-(4\pi\alpha A^{-1}/e^2)^2 \vert \boldsymbol\epsilon_c \cdot \textbf{R}_{eg} \vert^4 } + i (4\pi\alpha A^{-1}/e^2) \vert \boldsymbol\epsilon_c \cdot\textbf{R}_{eg} \vert^2 ]
$
and the polarizability tensor, defined by $ \textbf{R}_{\text{induced}}(\omega) = \boldsymbol\alpha(\omega) \cdot \textbf{E}_{\text{perturbation}}(\omega) $ \cite{ullrich2011}, as 
\begin{align*}
\Im \alpha_{\mu\nu}(\omega) &= \sum_{n=1}^\infty 2R^{(\mu)}_{gn}R^{(\nu)}_{ng} \frac{\Im \Omega_n}{(\hbar\omega-\Re \Omega_n)^2 + (\Im \Omega_n)^2}\\
&= 2R^{(\mu)}_{ge}R^{(\nu)}_{eg} \frac{\hbar\Gamma_{rr}}{(\hbar\omega-\Re \Omega_n)^2 + \hbar^2\Gamma_{rr}^2}
\end{align*}
with photoabsorption cross-section and linewidth
\begin{align*}
\sigma(\omega) &= \frac{4\pi\omega}{c}\Im \alpha(\omega),\quad
\Gamma_{rr} &=  \frac{4\pi\alpha\omega_{eg}}{e^2}A^{-1} \vert \boldsymbol\epsilon_c \cdot\textbf{R}_{eg} \vert^2~.
\end{align*}
A system has therefore no longer discrete excitations as common in TDDFT but features a physical linewidth $\Gamma_{rr}$. In the perturbative limit, the latter is identical to the Wigner-Weisskopf solution $\Gamma^{1D}_{WW} = \omega_{eg} \vert \boldsymbol\epsilon_c \cdot\textbf{R}_{eg} \vert^2A^{-1}/\hbar \varepsilon_0 c = \Gamma_{rr}$ (see SI). 

The radiative-reaction leads to a slight shift of the resonance as a consequence of the electromagnetic recoil $ \propto 1-(4\pi\alpha A^{-1}/e^2 \vert \boldsymbol\epsilon_c \cdot \textbf{R}_{eg} \vert^2)^2 /2 $. This Lamb-shift effect indicates that the mass of the particle is changed, resulting in an adjusted physical mass of the electron. 
For Purcell-enhanced emission in a waveguide, the Lamb-shift can take non-negligible values, especially when collective effects lead to collective Lamb-shift effects \cite{rohlsberger2010collective}. We obtain here for hydrogen $\approx 7$~meV with $A^{-1}=1/a_0^2$. However, for the vast majority of applications this shift is negligible and aspects such as the quality of the exchange-correlation potential in TDDFT are far more influential.

Let us point out that clearly also \textit{ab initio} QED and QEDFT benefit from the radiation-reaction approach. Typically, only a few cavity modes play a substantial role in the strong coupling and the manifold of weakly interacting modes describe the emission profile, i.e., the loss-channels, of the cavity. Using a radiation-reaction potential which represents a bath of free-space modes, cavity losses into free-space can be efficiently modeled as we will see in the following.
Such a separation is of imminent importance for the future development of \textit{ab initio} QED as it allows us to utilize higher level descriptions for a few most relevant cavity modes which explicitly account for the quantum features of light (see e.g. \cite{schafer2021making}) while retaining the full manifold of modes which account on a simplified level for emission, linewidth and loss. In this sense, the illustrated potential acts as highly efficient realization of a bath, comparable to open-system strategies in quantum optics.

\begin{figure}[t!]
\includegraphics[width=1.0\columnwidth]{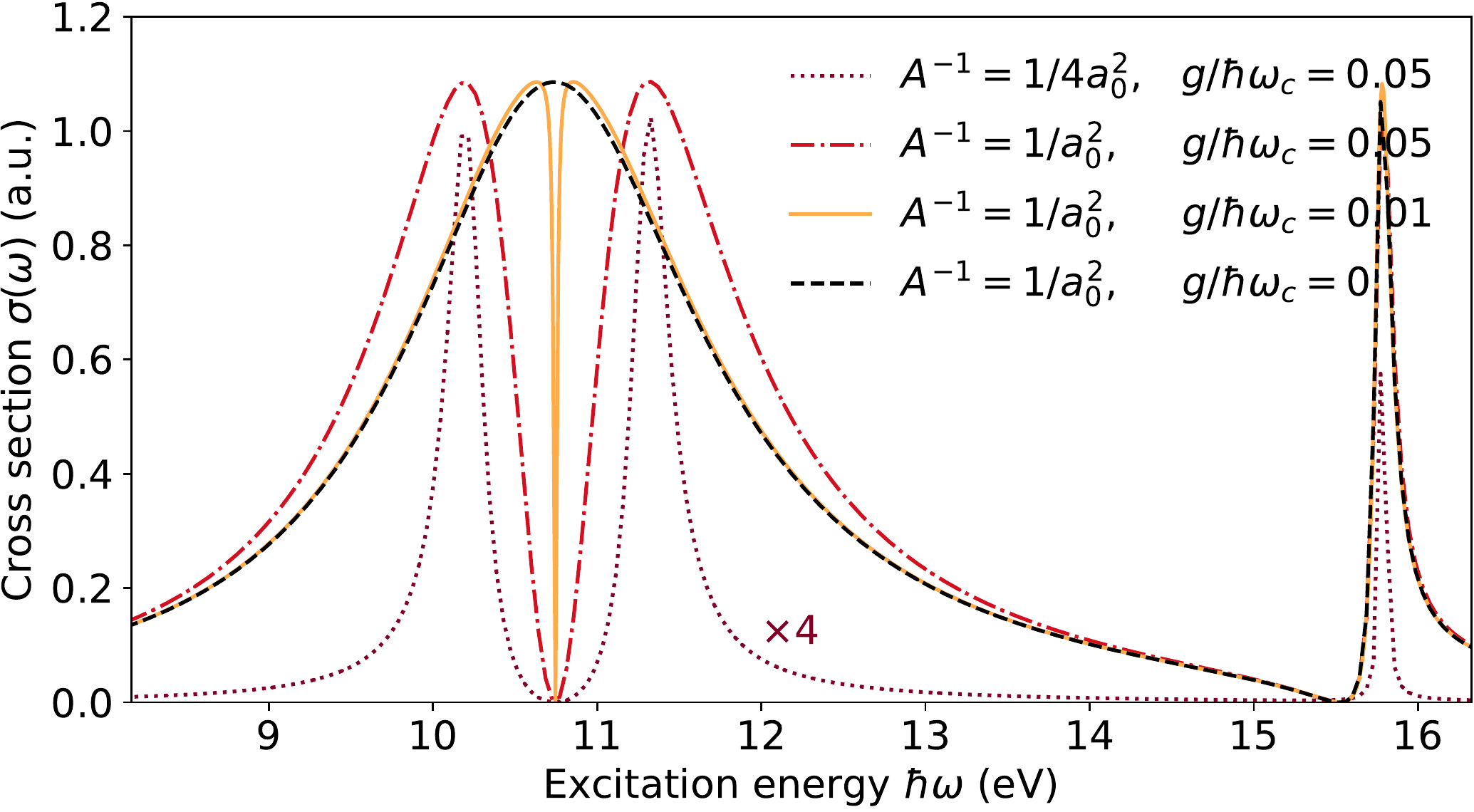}
\caption{ 
Photoabsorption cross-section $\sigma(\omega)$ of one-dimensional hydrogen coupled to a single cavity mode (strength $g=ea_0\sqrt{\hbar\omega_c/2\varepsilon_0V}$) in resonance to $\Omega_{eg}=10.746$~eV. With increasing strength of the photon-bath $A^{-1}$, i.e., decreasing quality of the hydrogenic oscillator, the system moves from slightly broadened polaritonic resonances into the regime of electromagnetically induced transparency. Numerical details in SI.}
\label{fig:hydrogen1d_cavityplusfriction}
\end{figure}

\textit{Electromagnetically induced transparency - }
One excellent example of the strength of such an photonic-bath treatment is the description of electromagnetically induced transparency (EIT) \cite{fleischhauer2005electromagnetically,peng2014and}. Let us couple a single loss-free cavity mode, 
representing for instance a whispering-gallery mode, strongly to the hydrogen system. 
In addition, the matter system can emit light into a waveguide, i.e., a bath of photonic modes according to eq.~\eqref{eq:rrpot1d}. Fig.~\ref{fig:hydrogen1d_cavityplusfriction} presents how the absorption-profile of hydrogen changes with increasing strength of the bath and the coupling strength to the single mode. The interplay between the high-quality cavity mode and the lossy electromagnetic bath induces a window of transparency for strong radiative emission, i.e., the natural linewidth (black, dashed) obtains a sharp window of transparency (yellow, solid) at which the system can no longer absorb light. For increasing $g/\hbar\omega_c$, hydrogen and single mode begin to hybridize, resulting in the polaritonic states that characterize strong light-matter coupling. EIT has a plethora of technological applications, including effectively stopping or storing light \cite{hau1999light,liu2001observation}, which become now for the first time available to TDDFT and QEDFT in a simple and intuitive way. 

While it is common practice to describe superradiant effects with quantum optical models \cite{gross1982superradiance} and the Purcell enhancement of spontaneous emission as perturbative correction to Wigner-Weisskopf theory \cite{koenderink2010use}, we show here that the radiation-reaction potential equips \textit{ab initio} frameworks with the necessary tools to address those aspects self-consistently. In order to describe realistic systems, we implemented the radiative-reaction potential introduced in eq.~\eqref{eq:rrpot1d} in the TDDFT code GPAW \cite{enkovaara2010electronic} and use the computationally efficient LCAO basis \cite{kuisma2015localized} (see \cite{gpawrrpotential} for a tutorial).

\begin{figure}
\includegraphics[width=1.0\columnwidth]{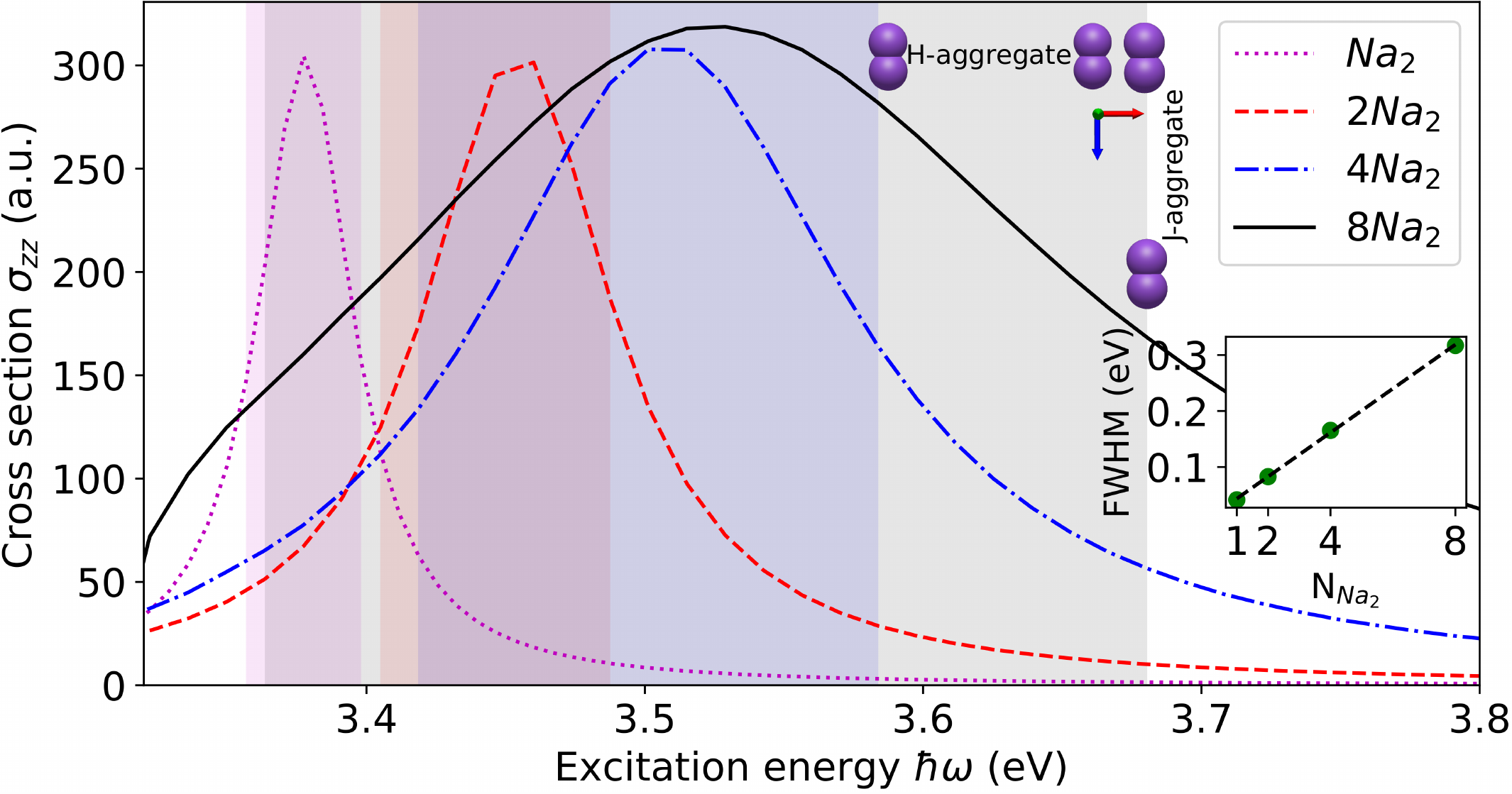}
\caption{Photoabsorption cross-section $\sigma_{zz}(\omega)$ (zz-component) for chains with variable length of far separated Na$_2$. The dimer axis is oriented along z with a bond distance of $1.104$~\AA, the chain along x has separations of $8$~\AA~ (H-aggregate configuration). We used eq.~\eqref{eq:rrpot1d} with a quantization area of $35.05$~\AA$^2$ 
and polarization along z. 
The emission rate/linewidth of the excitation around $3.5$~eV increases linearly $\Gamma \approx N_{Na_2} \Gamma_{Na_2}$ with the length of the chain (see inset, FWHM$=2\Gamma$ highlighted) as expected for the single-photon superradiant emission. H- and J-aggregate configuration are schematically illustrated. Numerical details in SI.}
\label{fig:superradiant}
\end{figure}

\textit{Superradiant linewidth - }
The synchronized evolution among a set of atoms or molecules amplifies their interaction  with the photonic environment. Superradiance describes the effect that the synchronized emission is quicker than the individual emission \cite{gross1982superradiance}. For the single-photon absorption spectrum this takes the form of a linearly increasing  linewidth with the size of the ensemble ($\Gamma \approx N_{Na_2} \Gamma_{Na_2}$) and is illustrated in fig.~\ref{fig:superradiant} for a set of Sodium dimers.
The decisive strength of the \textit{ab initio} implementation is the consistent treatment of radiative emission and matter, i.e., any form of Coulomb mediated interaction or charge-migration is treated self-consistently. 
Fig.~\ref{fig:superradiant} illustrates for instance a blue-shift that originates from the Coulomb-mediated dipolar coupling between the individual dimers as observed in H-aggregates \cite{kasha1965exciton}. Arranging the dimers in a head-to-tail orientation produces a red-shift as typical for J-aggregates. The radiation-reaction potential provides therefore consistently access to typical quantum optical and quantum chemical observables.

\begin{figure}[ht!]
\includegraphics[width=1.0\columnwidth]{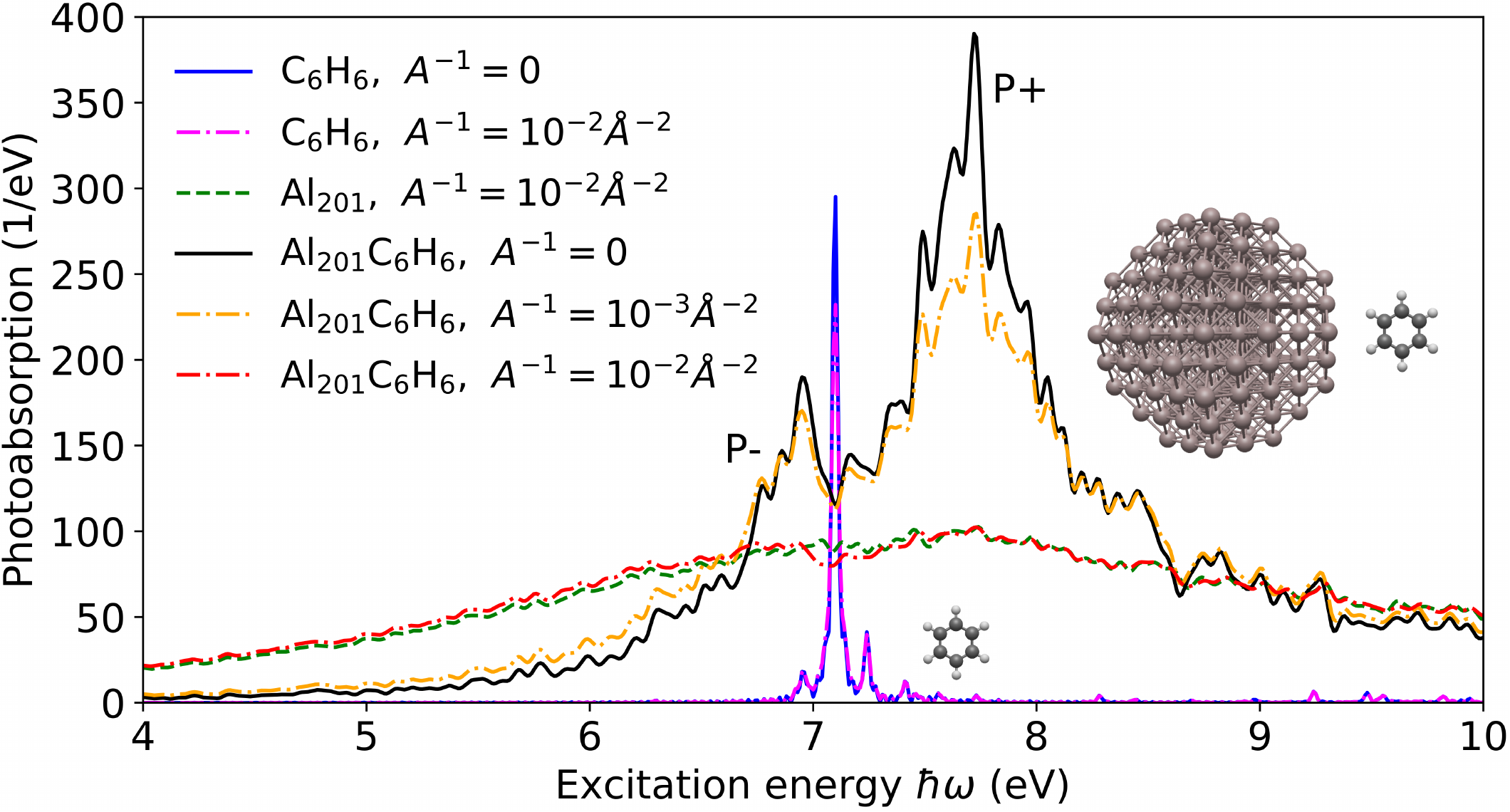}
\caption{Absorption spectrum in x-direction for isolated benzene (blue, magenta) and benzene strongly coupled to Al$_{201}$ (black, red, yellow - upper and lower hybrid state P$\pm$ indicated). Dashed(-dotted) lines include emission via the radiation-reaction potential with polarization $\boldsymbol\epsilon_c = \textbf{e}_x$. The sharp excitation of benzene is only marginally broadened when including emission (magenta, dashed-dotted), i.e., atoms and molecules have comparably long lifetimes. The localized surface-plasmon of the Al$_{201}$ cluster radiates on the other hand substantially stronger (green, dashed). Coupled benzene inherits the short lifetime of the plasmon (red/yellow dashed-dotted) -- any excitation is quickly transferred into the plasmon and either internally dephased or subsequently radiated into free-space. Numerical details in SI}
\label{fig:purcell}
\end{figure}

\textit{Plasmonic Purcell-enhancement - }
Plasmonic particles contribute in two ways to the quick decay of nearby molecules. First, strong dipolar oscillations of localized surface plasmons result in quick radiative decay. Second and often dominant, they feature very quick internal dephasing on the fs timescale due to Landau damping and electron-electron scatterings. In combination with charge migration and hot-electron transfer between nanoparticle and molecule, this results in a complex dynamic which is theoretically challenging to capture. Full-fledged \textit{ab initio} approaches provide here valuable results and following the radiation-reaction, we can now easily and consistently account for radiative features which enriches quantum optical and electronic structure perspectives by a linking framework.
Fig \ref{fig:purcell} illustrates 
the dipolar spectrum of benzene next to Al$_{201}$ \cite{rossi2019strong} including and excluding the radiative-emission. The strong longitudinal fields around the plasmon lead to a hybridization of the bare plasmonic and $\pi-\pi^*$ benzene excitation, a purely Coulombic feature. For small cross-sectional surfaces $A=10^2$ \AA$^2$, the radiative contributions increase in relevance and strong plasmonic currents provide the previously slow radiating benzene excitation with an efficient emission channel -- Purcell-enhancement $\Gamma_{Al_{201}C_6H_6}/\Gamma_{C_6 H_6}=114.5$. The larger $A$ the stronger internal dephasings compete with radiation, up to the point ($A^{-1}<10^{-3}$\AA$^{-2}$) where the Purcell-enhancement for the emission of benzene is dominated by those internal losses.
%
The competition between radiative and non-radiative decay channels is often tilted in favor of  internal losses while nanoantenna design can invert this characteristic \cite{rogobete2007design,baumberg2019extreme}.

\textit{Conclusion - }
Utilizing the dyadic Green's tensor and deriving a subsequent local radiation-reaction potential, we illustrated how the electromagnetic fields can be easily embedded into electronic structure theory.
This simple and computationally efficient ansatz allows to describe self-consistent light-matter interaction from first-principles and is equivalent to solving Maxwell and Schr\"odinger equation hand in hand.
The introduced radiation-reaction forces represent the recoil that ensures energy conservation during the emission of light.
We demonstrated this ansatz for a simplified one-dimensional waveguide at the example of TDDFT, providing radiative emission,  natural linewidths, strong-coupling, electromagnetically induced transparency, Purcell enhancement and superradiant emission with virtually no additional computational effort.
While we avoid the need to keep track of the electromagnetic fields, they are determined at any point in time by the dyadic Green's tensor and the electronic currents, allowing a precise description of the experimentally measurable fields. Our ansatz is of generic use for any time-dependent electronic structure theory and can be extended to the nuclear degrees of freedom. In addition, we present generalizations to more complex electromagnetic environments in the SI. 
Implementing the radiation-reaction potential into the large-scale TDDFT code GPAW illustrates the high accessibility and seamless integration of our approach into existing \textit{ab initio} libraries, paving the way for a stronger integration of light-matter and open-system features into the popular electronic structure approaches.
A separation into microscopic $\textbf{j}$ and macroscopic components $\varepsilon_r$ and the subsequent embedding of those macroscopic contributions into $\hat{V}_{rr}$ paves a way to conveniently describe collective strong coupling at the microscopic level. Collective strong coupling obtained recent interest in QED chemistry \cite{ebbesen2016} and its application will be detailed in a forthcoming publication.

\begin{acknowledgments}
We thank Jakub Fojt and Tuomas Rossi for assistance with GPAW as well as Michael Ruggenthaler and Ilya Tokatly for fruitful discussions. This work was supported by the Swedish Research Council (VR) through Grant No. 2016-06059 
and the computational resources provided by
the Swedish National Infrastructure for Computing (location Ume\aa) partially
funded by the Swedish Research Council through grant agreement no. 2018-05973.
\end{acknowledgments}

\bibliography{sponemission} 

\end{document}


\title{Supplemental Information: A Shortcut to Self-Consistent Light-Matter Interaction and Realistic Spectra from First-Principles}

  \author{Christian Sch\"afer}
  \email[Electronic address:\;]{christian.schaefer.physics@gmail.com}
  \affiliation{Department of Microtechnology and Nanoscience, MC2, Chalmers University of Technology, 412 96 G\"oteborg, Sweden}
  \affiliation{Department of Physics, Chalmers University of Technology, 412 96 G\"oteborg, Sweden}  
  \author{G\"oran Johansson}
  \affiliation{Department of Microtechnology and Nanoscience, MC2, Chalmers University of Technology, 412 96 G\"oteborg, Sweden}
  \affiliation{Department of Physics, Chalmers University of Technology, 412 96 G\"oteborg, Sweden}  
  
\date{\today}

\maketitle

\section{Light-matter Hamiltonian in dipole approximation}

We provide here a short derivation of the dipolar light-matter Hamiltonian.
Starting with the free Maxwell-equations, we solve the wave-equation $\nabla^2 \textbf{A}(rt) - \frac{1}{c^2} \partial_t^2 \textbf{A}(rt) = 0$ implying the transversality condition of the Coulomb gauge $ \nabla \cdot \textbf{A}(r) = 0 $. Notice that the Coulomb gauge should be chosen as otherwise the Coulomb interaction takes a different form and thus all electronic structure theory includes the wrong local description. This is a natural limitation for all QED quantization strategies and becomes problematic when quantizing for instance dissipative partially longitudinal+transversal modes such as in macroscopic QED. Certainly the ideal situation would be to counteract any gauge-choice for the quantized modes by adjusting the electronic structure calculations but given the extend of electronic structure literature, this is a tedious task. It is therefore essential to only quantize the transverse modes in the following. Only transversal modes can emit into free space and contribute to the far-field radiation while longitudinal components decay quickly \cite{craig1998}.

A rectangular box $V=\prod_{i=1}^{dim} L_i$ with perfect conductor boundary condition $\textbf{n} \times \textbf{E}_\perp (\br) \propto \textbf{n} \times \textbf{A}(\br) = 0$ delivers for example transverse eigenmodes of the form \cite{power1982quantum}
\begin{align}\label{eq:eigenmode}
	\mathbf{S}_{\textbf{k}\lambda}(\br) &= \sqrt{\frac{2^3}{V}}
	\begin{pmatrix}
	\epsilon_{\textbf{k}\lambda}^{(x)} \cos(k_x x) \sin(k_y y) \sin(k_z z) \\
	\epsilon_{\textbf{k}\lambda}^{(y)} \sin(k_x x) \cos(k_y y) \sin(k_z z) \\
	\epsilon_{\textbf{k}\lambda}^{(z)} \sin(k_x x) \sin(k_y y) \cos(k_z z)
	\end{pmatrix}
\end{align}
with momentum $\textbf{k} = (k_x, k_y, k_z)^T$, polarization-index $\lambda$ for the polarization vector $\boldsymbol\epsilon_{\textbf{k}\lambda}$ and free-field dispersion from the Helmholtz-equation $\omega(\textbf{k}) = c\vert \textbf{k}\vert$. The photonic field coordinates are represented by the harmonic oscillator coordinate $q_{\textbf{k}\lambda}$.
Electric and magnetic fields are given by 
\begin{align*}
\mathbf{A}(\br,t) &= \sum_{\textbf{k}\lambda} \sqrt{\frac{c^2}{\varepsilon_0}} q_{\textbf{k}\lambda}(t) \mathbf{S}_{\textbf{k}\lambda}(\br)\\
\mathbf{E}_{\perp}(\br,t) &= - \frac{1}{c}\partial_t \mathbf{A}(\br,t)\\
\mathbf{B}(\br,t) &=  \frac{1}{c}\mathbf{\nabla}\times \mathbf{A}(\br,t).
\end{align*}
and construct the transverse electromagnetic field energy
\begin{align*}
E_{\perp} = \frac{\epsilon_0}{2} \int d\textbf{r} \mathbf{E}_{\perp}(\br,t)^2 + c^2\mathbf{B}(\br,t)^2~.
\end{align*}

In the next step, we promote the canonical displacement and momentum coordinates to operators
\begin{align*}
\hat{q}_{k} &= \left(\frac{\hbar}{2\omega_{k}}\right)^{1/2} \left(\hat{a}^\dagger_k + \hat{a}_k\right)\\
\hat{p}_{k} &= i\left(\frac{\hbar \omega_{k}}{2}\right)^{1/2} \left(\hat{a}^\dagger_k - \hat{a}_k\right)
\end{align*}
with $[\hat{a}_i,\hat{a}_j^\dagger] =  \delta_{ij}$ such that 
\begin{align*}
	\hat{\textbf{A}}(\br) &= \sum_{\textbf{k}\lambda} \sqrt{\frac{c^2}{\varepsilon_0}} \hat{q}_{\textbf{k}\lambda} \mathbf{S}_{\textbf{k}\lambda}(\br)\\
	\hat{\textbf{E}}_{\perp}(\br) &= -\sum_{\textbf{k}\lambda} \sqrt{\frac{1}{\varepsilon_0}} \hat{p}_{\textbf{k}\lambda} \mathbf{S}_{\textbf{k}\lambda}(\br)\\
	\hat{\textbf{B}}(\br) &= \sum_{\textbf{k}\lambda} \sqrt{\frac{1}{\varepsilon_0}} \hat{q}_{\textbf{k}\lambda} \nabla \times \mathbf{S}_{\textbf{k}\lambda}(\br)~.
\end{align*}
The full minimal-coupling light-matter Hamiltonian in Coulomb-gauge reads then (for nuclear and electronic coordinates $q_i \in \{+eZ_i,-e\}$, $m_i \in \{M_n,m_e\}$, $\hat{p}_i \in \{-i\hbar\nabla_{N_{n_i}},-i\hbar\nabla_{N_{e_i}}\}$)
\begin{align*}
\hat{H}_{lm} &= \sum_{i=1}^{N_e+N_n} \frac{1}{2m_i} \left( \hat{p}_i - \frac{q_i}{c} \hat{\textbf{A}}(r)\right)^2  + \hat{H}_{ph} + \hat{V}_{ext}(\textbf{r},\textbf{R}_n) \notag \\ 
&+ \hat{W}_{ee}(\textbf{r},\textbf{r}') + \hat{W}_{nn}(\textbf{R}_n,\textbf{R}_n') + \hat{W}_{ne}(\textbf{r},\textbf{R}_n)\\
\hat{H}_{ph} &= \frac{\epsilon_0}{2} \int d\textbf{r} \hat{\mathbf{E}}_{\perp}(\br)^2 + c^2\hat{\mathbf{B}}(\br)^2 \notag\\ 
&= \frac{1}{2}\sum_{\textbf{k}} \sum_{\lambda=1}^{2} \hat{p}^2_{\textbf{k}\lambda} + \omega(\textbf{k})^2 \hat{q}^2_{\textbf{k}\lambda}~.
\end{align*}
We assume now that in the coupling-part around the domain over which the matter-system is expanded, the eigenmodes eq.~\eqref{eq:eigenmode} remain constant $\mathbf{S}_{\textbf{k}\lambda}(\textbf{r}) \approx  \mathbf{S}_{\textbf{k}\lambda}(\textbf{r}_0)$. Next, we employ the unitary Power-Zienau-Wooley transformation \cite{craig1998,schafer2019relevance} with the total dipole $\hat{\textbf{R}} = -\sum_{j=1}^{N_e}e\hat{\textbf{r}}_j +\sum_{j=1}^{N_n} eZ_j\hat{\textbf{R}}_j^{n}$. This transformation is nothing else than a momentum-translation which will transfer the coupling from momentum-fluctuations into displacement-fluctuations \cite{schafer2018insights}. By doing so, we remove the photonic part from the covariant momentum $\left( \hat{p}_i/m_i - q_i/c \hat{\textbf{A}}(\textbf{r}_{matter})\right) \rightarrow \hat{p}_i/m_i$ but introduce coupling in the previously pure photonic Hamiltonian $\hat{H}_{ph} \rightarrow \hat{H}_{ep}'$.
Performing a Fourier-transformation in the photonic coordinates $\hat{p}_{\textbf{k}\lambda} \rightarrow -\omega(\textbf{k})\hat{q}_{\textbf{k}\lambda}$, $\hat{q}_{\textbf{k}\lambda} \rightarrow 1/\omega(\textbf{k})\hat{p}_{\textbf{k}\lambda}$ leads finally to
\begin{align}\label{eq:hamiltonian}
\begin{split}
\hat{H} &= \hat{T} + \hat{V} + \hat{W} +  \hat{H}_{ep} \\
\hat{H}_{ep} &= \frac{1}{2} \sum_{\textbf{k}} \sum_{\lambda=1}^{2} \hat{p}_{\textbf{k}\lambda}^2 + \omega(\textbf{k})^2 \left[\hat{q}_{\textbf{k}\lambda}- \frac{\boldsymbol \lambda_{\textbf{k}\lambda}(\textbf{r}_0)}{\omega(\textbf{k})} \cdot \hat{\textbf{R}} \right]^2
\end{split}
\end{align}
with
\begin{align*}
\boldsymbol\lambda_{\textbf{k}\lambda}(\textbf{r}) = \frac{1}{\sqrt{\epsilon_0}} \mathbf{S}_{\textbf{k}\lambda}(\br)~.
\end{align*}
Here $\textbf{r}_0$ should be ideally the center of charge of the \textit{whole} matter system.
The PZW transformation removes the diamagnetic current from $[\nabla^2+\frac{\omega^2}{c^2}]\textbf{A}=-c\mu_0(-e\textbf{j}_{para,\perp})$ and furthermore
changes the field quantities, i.e., the meaning of $\hat{a}$ changes. 
The conjugated momentum is now the displacement field
\begin{align*}
\hat{\textbf{D}}_\perp &= \epsilon_0\sum_{\textbf{k}\lambda} \boldsymbol\lambda_{\textbf{k}\lambda}
\omega(\textbf{k})\hat{q}_{\textbf{k}\lambda}
\end{align*}
and the electric field is no longer a purely photonic quantity but includes the transversal polarization
\begin{align*}
\varepsilon_0\hat{\mathbf{E}}_\perp &= ( \hat{\mathbf{D}}_\perp  - \hat{\mathbf{P}}_\perp )\\
\hat{\mathbf{P}}_\perp &=  \sum_{\textbf{k}\lambda} \varepsilon_0 \boldsymbol\lambda_{\textbf{k}\lambda} (\boldsymbol\lambda_{\textbf{k}\lambda}  \cdot \hat{\textbf{R}})
\end{align*}
where
\begin{align*}
- \frac{1}{c}\partial_t \hat{A} = \hat{\textbf{E}}
\end{align*}
remains.

We will disregard the coupling to the nuclei for the TDDFT derivation.

\section{Radiation-reaction for arbitrary photonic environments}

\begin{figure}
\includegraphics[width=1.0\columnwidth]{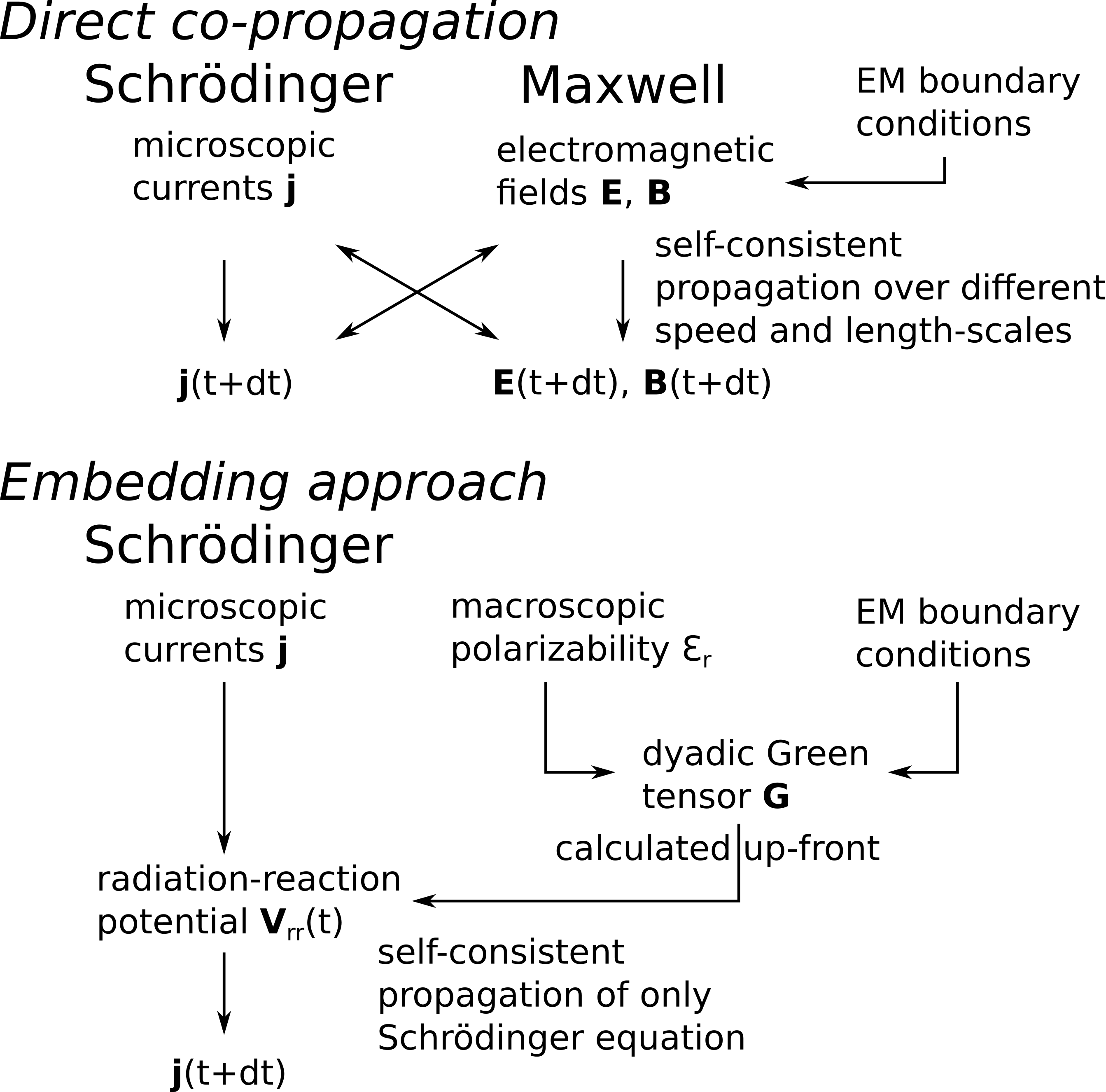}
\caption{Simplified schematic illustrating the advantage of the embedding ansatz via dyadic Green tensor and radiation-reaction potential. The electromagnetic fields are embedded into the Schr\"odinger equation via the potential which strongly simplifies the description of self-consistent interaction between light and matter from the matter perspective.}
\label{fig:concept}
\end{figure}

One-dimensional emission, as present in waveguides or idealized cavities, is surely only one out of many relevant realizations of electromagnetic environments. In general, the classical electromagnetic mode structure inside a given (lossy) resonator, which may include transversal and longitudinal components, can be obtained by solving the Helmholtz equation $\big[\nabla \times \frac{1}{\mu_r(\textbf{r}\omega)}\nabla \times - \omega^2 \mu_0\varepsilon_0\varepsilon_r(\textbf{r}\omega)\big] \textbf{G}(\textbf{r},\textbf{r}',\omega) = \boldsymbol\delta(\textbf{r}-\textbf{r}') $ with the boundary condition $\textbf{G}\rightarrow 0 ~\text{for}~\vert \textbf{r}-\textbf{r'}\vert \rightarrow \infty$, providing as solution the dyadic Green's tensor \cite{buhmann2013dispersionI}.
A generic electrodynamic environment couples in dipolar approximation to the electronic structure via the potential
$\hat{V}_{rr}(t)=-\hat{\textbf{R}}\cdot \textbf{E}_r(\textbf{r}_0 t) = -\hat{\textbf{R}}\cdot(\textbf{E}_{r,\parallel}(t) + \textbf{E}_{r, \perp}(t))$. We added the index $r$ only to clarify that we consider the fields generated by the matter-current, the longitudinal fields can not propagate into free-space. This is consistent with the classical limit of the dipolar PZW Hamiltonian eq.~\eqref{eq:hamiltonian}, i.e., 
$ \sum_{\textbf{k}\lambda}-\omega(\textbf{k}) \hat{q}_{\textbf{k}\lambda} (\boldsymbol\lambda_{\textbf{k}\lambda}  \cdot \hat{\textbf{R}}) + \frac{1}{2} (\boldsymbol\lambda_{\textbf{k}\lambda}  \cdot \hat{\textbf{R}})^2 \approx - \frac{1}{\varepsilon_0}\hat{\textbf{R}} \cdot \langle \hat{\textbf{D}}_\perp \rangle + 2\frac{1}{2\varepsilon_0}\hat{\textbf{R}} \cdot \langle \hat{\mathbf{P}}_\perp \rangle = -\hat{\textbf{R}}\cdot \textbf{E}_\perp$ and in addition the longitudinal Coulomb interaction.
Using $\textbf{E}_r(\textbf{r},\omega) = i\mu_0\omega\int_V dr' \textbf{G}(\textbf{r},\textbf{r}',\omega) \cdot (-e\textbf{j}(\textbf{r}',\omega))$, evaluated at the center of charge  $\textbf{G}(\textbf{r}_0,\textbf{r}_0,\omega) = \textbf{G}(\omega)$  and combined with a Fourier-transformation delivers
\begin{align}
\begin{split}
\hat{V}_{rr}(t)=&-\hat{\textbf{R}}\cdot \big[\mathcal{F}_t^{-1}(i\mu_0 \omega \textbf{G}(\omega)) \ast \int dr(-e\textbf{j}(\textbf{r},t) \big]\\
=&-\hat{\textbf{R}}\cdot \big[\mathcal{F}_t^{-1}(i\mu_0 \omega \textbf{G}(\omega)) \ast \dot{\textbf{R}}(t)\big]~.
\end{split}
\end{align}
Once $\textbf{G}(\omega)$ or its transverse projection $\textbf{G}_\perp(\omega)$ is obtained, e.g., using one of the widely used finite-difference frequency or time-domain solvers of Maxwell's equations, the necessary convolution can be effectively approximated by finite-difference methods. This embedding scheme ensures that electronic structure calculations remain computationally efficient while describing at the same time realistic electromagnetic environments. The conceptual difference between direct co-propagation of Maxwell and Schr\"odinger equation and the embedding approach is illustrated in fig.~\ref{fig:concept}.

All applications in this work will focus on analytic solutions for the transversal electromagnetic environments, i.e., the coupling between transverse fields characterized by $\textbf{G}_\perp(\omega)$ and the from first-principles described material. Since dipole-moment (or respectively current) are routinely and cost-efficiently calculated observables for the first-principles approaches, the computational effort to include $\hat{V}_{rr}(t)$ is vanishingly small.
The description of small plasmonic systems, featuring localized surface plasmons, can be done most consistently by simply describing also the plasmonic system with TDDFT, i.e., on equal footing with any molecular particle close to it \cite{rossi2019strong}. Obvious computational limitations will prevent us from pushing this 'single-system' description to arbitrary system-size which motivates mixed quantum-classical descriptions such as combining Mie-theory and TDDFT via the radiation-reaction potential. Consider hereby the disclaimer at the start of this document regarding the consistency of gauges, i.e., electronic structure theory is usually defined in the Coulomb gauge and the radiation-reaction forces should be consistent with this choice.

Let us remind the reader that a quantization of the here introduced classical expressions is not permitted as it will miss the previously introduced self-polarization term $(\boldsymbol\lambda\cdot\hat{\textbf{R}})^2/2$. 
Quantization of quasi-normalmodes has been explored by macroscopic QED and associated approaches \cite{buhmann2013dispersionI,franke2019quantization,feist2021macroscopic}. The target of most of these techniques is to translate the complex mode-structure around sizable plasmonic systems into easy to handle (quasi) normal-modes which allows the usage of common quantum optical techniques to describe e.g. emission via the plasmonic modes. The number of those normal-modes will be considerable and commonly only perturbative usage along the lines of Wigner-Weisskopf theory is employed.
The combination of radiation-reaction and perturbation theory delivers comparable observations \cite{harel2000spontaneous}. 

Extensions beyond the dipolar form (PZW gauge and subsequent long-wavelength approximation) are possible but demand careful scrutinization of the involved currents and fields. A short discussion of a radiation-reaction vector-potential for extended system is discussed in sec.~\ref{sec:extended}. In particular, the electric dipole approximation fixes the physical current to be equivalent to the paramagnetic current. In full minimal Coupling, this does no longer hold and the physical current comprises paramagnetic, diamagnetic and potentially Stern-Gerlach-like contributions \cite{jestadt2018real,schafer2019relevance,schafer2021making}. Irrespective thereof, the vector-potential in Coulomb-gauge is the solution to the Helmholtz equation
$ [\nabla^2+\frac{\omega^2}{c^2}]\textbf{A}=-c\mu_0(-e\textbf{j}_\perp) $ with the transverse component of the physical current. The corresponding dyadic Green's tensor (from which we might isolate linear media) evidently provides then the minimally coupled fields depending on the transverse currents. Alternatively, explicit multi-polar expansions of the electric and magnetic fields could be coupled as both are determined by the same physical current and dyadic tensor $\textbf{B} = \frac{1}{i\omega}\nabla\times\textbf{E}$ \cite{buhmann2013dispersionI}. In any case, obtaining and using the full spatially resolved dyadic Green's tensor is computationally far more challenging. Extensions beyond the dipolar case would furthermore imply the usage of current-density functional-theory.
Entering the realm of current-DFT has however the advantage that usage of the Vignale-Kohn functional \cite{vignale1996current} provides access to non-adiabatic electron-electron interactions which improves the description of electronic relaxation processes, e.g., inside nanoplasmonic particles. Extending the here proposed embedding scheme to allow a consistent description of minimal coupling from first-principles is possible but certainly challenging.

\subsection{Single-dimensional time-dependent radiation-reaction potential via dyadic Greens function}

The following derivations will focus on the methodology of time-dependent density-functional theory but can be abstracted to the more generic case.
We start from the expressions $v_{rr}(\textbf{r}t)=e\textbf{r}\cdot\textbf{E}_{r,\perp}(t)$
and $\textbf{E}_r(\textbf{r},\omega) = i\mu_0\omega\int_V dr' \textbf{G}(\textbf{r},\textbf{r}',\omega) \cdot (-e\textbf{j}(\textbf{r}',\omega))$ using the Greens function as defined by the spectral theorem 
$\textbf{G}(x,x',\omega) = \sum_{\textbf{k}\lambda} \frac{S(x)S(x')}{k^2-(\omega/c)^2} \boldsymbol\epsilon_c \boldsymbol\epsilon_c^T$ with $S(x) = \sqrt{1/V}(\cos(2\pi x/L)+\sin(2\pi x/L))$.
The electric field can be grouped into two contributions that appear in frequency space
\begin{align*}
\frac{1}{k^2-\omega^2/c^2}\quad \text{and} \quad i\omega\textbf{j}(\textbf{r}\omega)~.
\end{align*}
In order to obtain the time-dependent radiation-reaction field, we make use of the convolution theorem for the inverse Fourier-transformation $\mathcal{F}^{-1}(\textbf{E}_r(\textbf{r},\omega)) \propto \mathcal{F}^{-1}( \frac{1}{k^2-\omega^2/c^2}  i\omega\textbf{j}(\textbf{r}\omega) ) = \mathcal{F}^{-1}( \frac{1}{k^2-\omega^2/c^2}) \ast \mathcal{F}^{-1}(i\omega\textbf{j}(\textbf{r}\omega) )$, with $\mathcal{F}^{-1} f(\omega) = \int \frac{d\omega}{2\pi} e^{-i\omega t} f(\omega)$. Then, $\mathcal{F}^{-1}(i\omega\textbf{j}(\textbf{r}\omega)) = -\partial_t \textbf{j}(\textbf{r}t)$ is trivially obtained. The first part is formally solved by complex integration around the two poles at $k \pm \omega/c$
\begin{align*}
&\int \frac{d\omega}{2\pi} e^{-i\omega t} \frac{1}{(k-\omega/c)(k+\omega/c)}\\ 
&= c^2 \frac{i}{2ck}(e^{ickt}-e^{-ickt}),~t > 0; ~\text{and 0 otherwise.}
\end{align*}
The causality $t>0$ is absorbed into the Heaviside function $\theta(t)$ and we introduce the sine function such that the convolution takes the form
\begin{align*}
\int_{-\infty}^{\infty} dt' \theta(t-t') c^2 \frac{\sin(ck(t-t'))}{ck} (-1)\partial_t \textbf{j}(\textbf{r}t)
\end{align*}
and by partial integration (assuming $\textbf{j}(\textbf{r}, -\infty)=0$)
\begin{align*}
\int_{-\infty}^{t} dt' c^2 \cos(ck(t-t')) \textbf{j}(\textbf{r}t')~.
\end{align*}
The negative infinity is hereby limited by the initial conditions (see e.g. the following section).
Our radiated field is then given by (free-space $ck=\omega,~\mu_0=1/(\varepsilon_0 c^2)$)
\begin{align*}
\textbf{E}_r(x,t) = \boldsymbol\epsilon_c  \int_V dr' \sum_{\textbf{k}} \frac{1}{\varepsilon_0}S(x)S(x')\\ 
\int_{-\infty}^{t} dt'\cos(ck(t-t'))  \boldsymbol\epsilon_c\cdot (-e)\textbf{j}(\textbf{r}'t')
\end{align*}
and the radiation-reaction potential at the center of charge $x=x'=x_0=0$ follows immediately
$v_{rr}(\textbf{r}t)=e\textbf{r}\cdot\textbf{E}_{r,\perp}(t) = e\textbf{r}\cdot \boldsymbol\epsilon_c  \sum_{\textbf{k}} \frac{1}{V\varepsilon_0}\int_{-\infty}^{t} dt'\cos(ck(t-t')) \int_V dr'  \boldsymbol\epsilon_c\cdot (-e)\textbf{j}(\textbf{r}'t')$.
With the help of the continuity equation $\dot{\textbf{R}} = \int dr -e\textbf{j}(\textbf{r}t)$ 
\footnote{$\dot{\textbf{R}}(t) = \int dr (-e\textbf{r}) \dot{\rho}(\textbf{r}t) = \int dr (-e\textbf{r}) (-\nabla\cdot \textbf{j}_s(\textbf{r}t)) = \int dr (-e\textbf{j}_s(\textbf{r}t))$, with the paramagnetic Kohn-Sham current-density $\textbf{j}_s=\hbar/2m i\sum_{i} (\phi_i^*\nabla \phi_i-\phi_i \nabla \phi_i^*)$.} 
we obtain the compact form
\begin{align*}
v_{rr}(\textbf{r}t)= e\textbf{r}\cdot \boldsymbol\epsilon_c  \sum_{\textbf{k}} \frac{1}{V\varepsilon_0}\int_{-\infty}^{t} dt'\cos(ck(t-t')) \boldsymbol\epsilon_c\cdot \dot{\textbf{R}}(t').
\end{align*}
As we will see in the following section, this form is identical to the Maxwell/Ehrenfest potential in QEDFT describing the interaction between a set of harmonic oscillators and the electronic system. Turning the sum over modes into an explicit integration results ultimately in the radiation-reaction potential eq.~\eqref{eq:rrpot1d} as shown in the main text.

\subsection{Single-dimensional free-space emission as analytic limit from QEDFT}

Starting with the seminal work by Tokatly \cite{tokatly2013}, it is possible to explicitly derive a local potential for the Ehrenfest interaction between a set of photonic modes and the electronic system in the long-wave approximation
\begin{align*}
v_M(\textbf{r}t) &= \sum_{\textbf{k}\lambda} \boldsymbol\lambda_{\textbf{k}\lambda} \cdot e\textbf{r} \big[ \omega_{\textbf{k}\lambda} q_{\textbf{k}\lambda}(t) - \boldsymbol\lambda_{\textbf{k}\lambda}\cdot \textbf{R}(t) \big]~.
\end{align*}
The mode-resolved Maxwell equations describing those eigenmodes
$\partial_t^2 q_{\textbf{k}\lambda}(t) + \omega_{\textbf{k}\lambda}^2 q_{\textbf{k}\lambda}(t) = \omega_{\textbf{k}\lambda} \boldsymbol\lambda_{\textbf{k}\lambda}\cdot \textbf{R}(t)$
can be solved with the help of the classical Greens function $\sin(\omega_{\textbf{k}\lambda} (t-t'))/\omega_{\textbf{k}\lambda}$. 
Let us assume that at $t=t_0$ no time-dynamic was present (in the interacting and the Kohn-Sham system!), i.e., $\partial_tq_{\textbf{k}\lambda}(t_0) = \partial_t \textbf{R}(t_0) = 0$. Then $q_{\textbf{k}\lambda}(t_0) = \boldsymbol\lambda_{\textbf{k}\lambda}\cdot \textbf{R}(t_0)/\omega_{\textbf{k}\lambda} $ and we obtain after partial integration
\begin{align*}
v_M(\textbf{r}t) &= \sum_{\textbf{k}\lambda} \boldsymbol\lambda_{\textbf{k}\lambda} \cdot e\textbf{r} \int_{t_0}^{t} dt' \cos(\omega_{\textbf{k}\lambda} (t-t'))  \boldsymbol\lambda_{\textbf{k}\lambda}\cdot \dot{\textbf{R}}(t')~.
\end{align*}
The self-polarization term $\sum_{\textbf{k}\lambda} - \boldsymbol\lambda_{\textbf{k}\lambda} \cdot e\textbf{r}~ \boldsymbol\lambda_{\textbf{k}\lambda}\cdot \textbf{R}(t)$ is consequentially part of the causality $\theta(t-t')$.

With $S(0) = \sqrt{1/V},~k_x=2\pi n_x/L_x,~n_x\in \mathbb{Z}$, $\sum_{\textbf{k}\lambda} = \frac{L_x}{2\pi} \int_{-\infty}^{\infty}dk$, performing the explicit integration and we obtain with $\frac{\sin(\omega(t-t'))}{\pi(t-t')}\overset{\omega\rightarrow\infty}{\rightarrow}=\delta(t-t')$
the radiation-reaction potential describing one-dimensional emission within TDDFT as defined in the main text
\begin{align}
\label{eq:rrpot1d}
v^{1D}_{rr}(\textbf{r}t) &= \frac{4\pi\hbar\alpha}{e^2} A^{-1}~ \boldsymbol\epsilon_c \cdot \dot{\textbf{R}}(t) ~\boldsymbol\epsilon_c \cdot (-e\textbf{r})~.
\end{align}

We will focus in the following on other analytic limits that provide a very intuitive understanding of the radiation-reaction potential. First, the presence of two mirrors will modulate the free-space emission which illustrates nicely the conceptual step that has to be performed in order to describe more general systems. Finally, we extend our derivations to full three-dimensional emission which allows us to connect radiation-reaction to its historical counterpart, the Abraham-Lorentz model.

\subsection{Edge-emission from a cavity}
Let us assume we describe a cavity with confined modes in the z direction while we allow for free-space emission in x direction and align the polarization of all modes along the y direction for simplicity ($S(x,z)=\sqrt{2/V}\sin(\pi z/L_z)[\cos(2\pi x/L_x)+\sin(2\pi x/L_x)]$).
We separate the sum of modes into those aligned along the z-axis (confined modes) and those that are propagating with a very small $k_z = \frac{\pi}{L_z}n_z$ component
\begin{align*}
\sum_{k} = \sum_{k_x \gg k_z} + \sum_{k_z \gg k_x}
\end{align*}
Focusing on the first term
\begin{align*}
\sum_{k_x \gg k_z} &= \frac{L_x}{2\pi} \frac{L_z}{\pi} \int_{-\infty}^{\infty} dk_x \int_{0}^{\infty}d_z= \frac{L_xL_z}{2\pi^2} \int_0^\infty dk k  \int_{-\phi_m}^{\phi_m} 
\end{align*}
where we introduced polar coordinates in the x-z plane and assumed that we only treat k's which are very close to the x-axis. 
Setting $x=0$ without loss of generality, we then have
\begin{align*}
&\sum_{k_x \gg k_z} \boldsymbol\lambda_{\textbf{k}} \boldsymbol\lambda_{\textbf{k}} \cos(\omega_{\textbf{k}}(t-t')) \\
&= \frac{L_x L_z}{\varepsilon_0 V\pi^2} \int_0^\infty dk k  \int_{-\phi_m}^{\phi_m}  \sin^2(k\sin(\phi)z_0)\cos(\omega_{\textbf{k}}(t-t'))
\end{align*}
which for small $\phi$ can be expanded as
\begin{align*}
&= \frac{1}{\varepsilon_0 \pi^2 L_y} \int_0^\infty dk k  \int_{-\phi_m}^{\phi_m} \sin^2(k \phi z_0)\cos(\omega_{\textbf{k}}(t-t'))
\end{align*}
Performing the integration over $\phi$ leads to
\begin{align*}
= \frac{1}{\varepsilon_0 \pi^2 L_y}& \int_0^\infty dk k \cos(\omega_{\textbf{k}}(t-t')) \bigg( \phi_m - \frac{\sin(2\phi_m k z_0)}{2k z_0} \bigg)
\end{align*}
Next, we have to find an expression for $\phi_m$. The angle $\phi = \arctan(k_z/k_x)$ is in our case approximately $\approx k_z/k_x$ and we know that for $\omega \rightarrow \infty,~\omega =ck \approx c k_x$. The limit we are therefore about to perform suggests that
\begin{align*}
\phi_m \approx \frac{k_z}{k_x} \approx c\frac{k_z}{\omega} = \frac{c\pi n_z}{\omega L_z}
\end{align*} 
Only for small $k_z/k_x$ and thus $n_z \in \mathbb{N} \approx 1$ does our approximation make sense and we therefore set $n_z=1$ from here on.
Since $\phi_m \propto \frac{1}{\omega}$ we can now obtain an analytic solution for the  $dk=\frac{d\omega}{c},~k=\frac{\omega}{c}$ integral
\begin{align*}
&= \frac{1}{\varepsilon_0 c \pi^2 L_y} \bigg( \frac{\pi}{L_z} - \frac{\sin(2\pi z_0/L_z)}{2 z_0} \bigg)\frac{\sin(\omega_{\textbf{k}}(t-t')) }{(t-t')}\vert_{\omega_{\textbf{k}}=0}^{\omega_{\textbf{k}}\rightarrow\infty}
\end{align*}
Which simplifies to 
\begin{align*}
&= \frac{1}{\varepsilon_0 c L_y L_z} \bigg( 1 - \frac{\sin(2\pi z_0/L_z)}{2\pi z_0/L_z} \bigg) \delta(t-t')
\end{align*}
The sine expression tends for $z_0 \rightarrow 0$ towards 1, i.e., next to the mirror we do not have any available mode-volume and the emission is quenched (destructive Purcell). However, if we place our system in the center of the cavity $z_0=L_z/2$, we obtain $\sin(\pi)=0$ and thus 'standard' emission along the x direction. Notice that $z_0=\frac{3L_z}{2 2},~\sin(3\pi/2)/(3\pi/2)\approx-0.2122$ is small but negative, i.e., weak amplifications can appear depending on the position along $z_0$. This behavior resembles the typical dipole emission next to a perfect conductor plate \cite{vos2009orientation}.

While this provides the expected emission-behavior of a dipole next to a plate, we have two plates, the second at distance $z_0=L_z$. Taking this into account, a symmetrized multiplicative form should be used similar to
\begin{align*}
\left(1 - \frac{\sin(2\pi z_0/L_z)}{2\pi z_0/L_z}\right)\left(1 - \frac{\sin(2\pi (1-z_0/L_z))}{2\pi (1-z_0/L_z)}\right)~.
\end{align*} 
Under those conditions and considering $1/\varepsilon_0 c = 4\pi\hbar\alpha/e^2$, we obtain the radiation-reaction potential for perpendicular emission from a cavity (edge emission from a cavity with polarization along y) as
\begin{align*}
v^{edge}_{rr}(\textbf{r}t) =& v^{1D}_{rr}(\textbf{r}t) \left(1 - \frac{\sin(2\pi z_0/L_z)}{2\pi z_0/L_z}\right) \\
&\cdot \left(1 - \frac{\sin(2\pi (1-z_0/L_z))}{2\pi (1-z_0/L_z)}\right)
\end{align*}
which is only slightly modified with respect to the pure free-space emission $v^{1D}_{rr}$.

\subsection{Three-dimensional free-space emission and its connection to the Abraham-Lorentz model}

With increasing dimensionality of the photonic environment, the density of photonic modes increases $\mathcal{O}(\rho_{ph}^{d-1})$. 
The dyadic Green's tensor provides a particularly elegant approach which is detailed in the following.
In 3D free-space $\Re\textbf{G}(\textbf{r},\textbf{r}',\omega) \propto \frac{1}{\vert \textbf{r}-\textbf{r}' \vert}$ will diverge at the point $\textbf{r}=\textbf{r}'=\textbf{r}_0$. In contrast, the imaginary part provides the self-dyadic \cite{buhmann2013dispersionI}
\begin{align*}
\Im\textbf{G}(\textbf{r},\textbf{r},\omega)=\frac{\omega}{6\pi c}\textbf{1}~.
\end{align*}
With $\textbf{G}(\textbf{r}_0,\omega)\approx i\Im\textbf{G}(\omega)$, $\textbf{E}_r(\omega)=\mu_0 \textbf{G}(\textbf{r}_0,\omega) \cdot i\omega (-e) \textbf{J}(\omega)$ and $-e\textbf{J}(t)=\partial_t \textbf{R}(t)$, we obtain
\begin{align*}
\hat{V}_{rr}(t) = \frac{-1}{6\pi\varepsilon_0c^3} \partial_t^3 \textbf{R}(t) \cdot \hat{\textbf{R}}
\end{align*}
which delivers precisely the radiation-reaction force of the Abraham-Lorentz model
\begin{align*}
F_{AL} = \frac{e^2}{6\pi\varepsilon_0 c^3} \dddot{x}(t)~.
\end{align*}
Such a radiation-reaction jerk $\dddot{x}$, and the associated classical Newton equations, can result in non-causal pre-acceleration features and runaway solutions.\\
Calculating the third time-derivative is numerically expensive, unstable and in general not advised. In the specific TDDFT case, we could express the second derivative of the density in terms of the forces exerted by the Kohn-Sham stress-tensor $\hat{T}_{ik}(\textbf{r}) = \frac{1}{2}[\partial_i\hat{\Psi}^\dagger(\textbf{r})\partial_k\hat{\Psi}(\textbf{r}) + \partial_k\hat{\Psi}^\dagger(\textbf{r})\partial_i\hat{\Psi}(\textbf{r}) - \frac{1}{2}\partial_i\partial_k\{\hat{\Psi}^\dagger(\textbf{r})\hat{\Psi}(\textbf{r})\}]$  
\begin{align*}
\partial^2 \rho(\textbf{r}t) = \nabla \cdot [ \rho(\textbf{r}t) \nabla v_s(\textbf{r}t) ] + \langle \sum_{i,k} \partial_i \partial_k \hat{T}_{ik}(\textbf{r}) \rangle
\end{align*}
which would circumvent the third derivative \cite{van1999mapping}.
The dependence of $\partial^2 \rho$ on the radiation-reaction contribution in $v_s$ includes now the self-consistency that was previously encoded in the higher derivative. In a first intuitive step, one could discard the self-consistency of the radiation-reaction potential, i.e., $v_s = v_{ext}+v_{Hxc}$, but the detailed analysis necessary to judge the quality (including the question of bijectivity  between potential and density) of the suggested approximations extends beyond the scope of this work.\\
Alternatively, we can circumvent the problematic high-order time-derivative by using a Markovian-like approximation for the second derivative based on the harmonic assumption $\ddot{\textbf{R}} + \omega_n^2 \textbf{R} = 0$.
Then,
\begin{align*}
\hat{V}_{rr}(t) = \frac{-\omega_n^2}{6\pi\varepsilon_0c^3} \partial_t \textbf{R}(t) \cdot \hat{\textbf{R}}
\end{align*}
and following the linear response derivation, the natural linewidth
\begin{align*}
\Gamma_{rr}^{3D} = \frac{\omega_n^3 \vert \textbf{R}_{n} \vert^2 }{6\hbar\varepsilon_0 \pi c^3}
\end{align*}
is identical to the free-space Wigner-Weisskopf rate $\Gamma_{WW} = \frac{\omega_n^3 \vert \textbf{R}_{n} \vert^2 }{3\hbar\varepsilon_0 \pi c^3}$ up to a factor 2 that originates from the two possible polarizations into which the oscillating charge can radiate.

\subsection{Radiation-reaction vector-potential for the application to extended systems}\label{sec:extended}

If we intend to investigate extended systems, a reformulation into the form of a radiation-reaction vector-potential would be preferred as the dipolar operator is not well defined with periodic boundaries.
The start is rather simple. With $\textbf{E}_\perp = -\frac{1}{c}\partial_t \textbf{A}$ we obtain
$ \textbf{A}(\textbf{r},\omega) = \mu_0 c \int dr' \textbf{G}_\perp (\textbf{r},\textbf{r}', \omega) (-e)\textbf{j}_{physical}(\textbf{r}'\omega) $ which we can use in the coupling Hamiltonian in long wavelength approximation
$\hat{H}_{LM} = \frac{e}{c}\textbf{A}(t) \cdot \hat{\textbf{J}}_{physical},~\hat{\textbf{J}}_{physical} = \frac{-i\hbar}{m_e}\sum_i \nabla_i - \hat{\textbf{J}}_{dia}$.
Unfortunately, the diamagnetic current complicates this procedure as it leads to nested dependencies. If we represent $\textbf{G}$ in a sum of harmonics, it is possible to absorb the diamagnetic contributions into redefined frequencies $\omega \rightarrow \tilde{\omega},~\tilde{\omega}^2 = \omega^2 + \frac{N}{\varepsilon_0 V}$ and polarizations $tilde{\boldsymbol\epsilon}$ with the help of a Bogoliubov transformation (see e.g. \cite{schafer2021making}). The poles of $\textbf{G}$ move then to $\tilde{k}=\tilde{\omega}/c$ and one obtains for the example of the one-dimensional waveguide
$\textbf{A}_r(t) = \sum_{k} \frac{c \boldsymbol\epsilon_c \boldsymbol\epsilon_c^T}{\varepsilon_0 V} \int_{-\infty}^t dt' \frac{\sin(c\tilde{k} (t-t'))}{c\tilde{k}} \textbf{J}_{para}(t') $. The analytic integration of this expression is however challenging. If the involved fields remain weak enough such that the diamagnetic contributions (which are of order $\mathcal{O}(\alpha^2)$) remain small, we can approximate $\tilde{k}\approx k$ and perform with $\int_{-\infty}^\infty \frac{d\omega}{c} \frac{\sin(\omega (t-t'))}{\omega} = \frac{\pi}{c} ~\forall ~t-t' > 0$ the integration to obtain
$\textbf{A}_r(t)= \frac{\boldsymbol\epsilon_c \boldsymbol\epsilon_c^T}{\varepsilon_0 A} \int_{-\infty}^t dt' \textbf{J}_{para}(t')$. We then obtain the corresponding radiation-reaction expression $\hat{H}_{LM}\approx \frac{e}{c} (\textbf{A}_{drive}(t) + \frac{\boldsymbol\epsilon_c \boldsymbol\epsilon_c^T}{\varepsilon_0 A} \int_{-\infty}^t dt' \textbf{J}_{para}(t')) \cdot \hat{\textbf{J}}_{para}$.

\section{Strong field effects using the radiation-reaction approach}
\begin{figure}
\includegraphics[width=1.0\columnwidth]{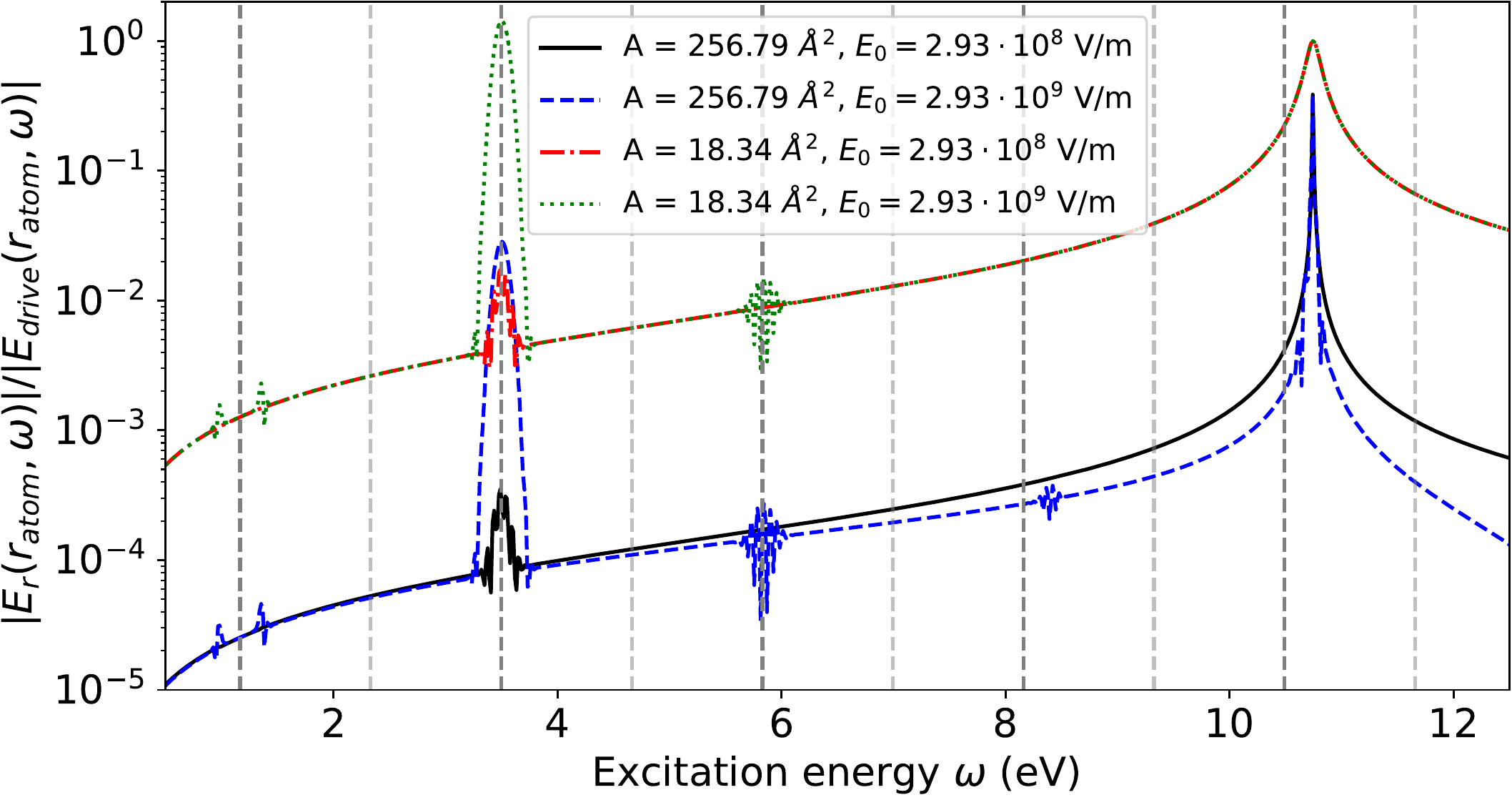}
\caption{HHG spectrum for one-dimensional hydrogen driven by a laser-pulse $E_{drive}(r_{atom},t)=E_0 \sin(\omega_L t) exp\{-\frac{(t-t_0)^2}{\sigma_t^2}\}, ~\omega_L=1.166 eV,~t_0=72.57~fs,~\sigma_t = 24.19~fs$. The radiated field at the atomic position $E_r(r_{atom},\omega)$ is obtained from the 1d-waveguide radiation-reaction potential as introduced in the main text. Vertical gray-dashed lines indicate the harmonics. We describe the electronic subsystem on 601 grid points with $\Delta x = 0.05~a.u.$. The combined system is propagated for $t_{max}=12000~a.u.$ with a time-step $\Delta t=5 \cdot 10^{-3}~a.u.$.}
\label{fig:hhgspectrum}
\end{figure}

Embedding the electromagnetic fields into the local radiation-reaction potential is not limited in any sense by the strength of the involved fields as long as we treat all materials explicitly as microscopic currents $\textbf{J}(t)$. Clearly, this changes when we absorb parts of the microscopic material into effective macroscopic linear media via $\varepsilon_r(\textbf{r}\omega),~\mu_r(\textbf{r}\omega)$. In other words, the radiation-reaction ansatz is only limited in the involved field-strengths if $\textbf{G}$ is the solution to the Helmholtz equation that involves linear media. Its limits are the same as solving Maxwells equations and electronic structure (e.g. via TDDFT) hand in hand.
Our example of the one-dimensional waveguide for example is valid for arbitrary field strength. Fig.~\ref{fig:hhgspectrum} illustrates the high-harmonic generation (HHG) of radiated fields $\textbf{E}_r(r_{atom},t)=-\frac{4\pi\hbar \alpha}{e^2 A}\boldsymbol \epsilon_c ~\boldsymbol \epsilon_c \cdot \dot{\textbf{R}}(t)$ by driving one-dimensional hydrogen in the nonlinear regime. Smaller cross-sectional areas of the waveguide result in overall stronger emission (where the recoil-damping can overshadow higher harmonics) but the strength of the HHG signal scales foremost with the intensity of the driving field. Interestingly, while we observe the expected odd harmonics, closer to the first excitation energy of hydrogen a small blue-shift is visible.

\section{Numerical Details}
Inputs and code are available upon reasonable request. Multiplicative factor indicate the value with which the shown curve has to be multiplied in order to obtain the calculated data. In order to not interfere with the linear response kick, the radiation-reaction potential has been switched on always shortly after the kick (hydrogen runs at t=2~a.u., GPAW at t=5~a.u.).
Fig.~1 and 2 utilized a time-stepping of $\Delta t=10^{-2}~a.u.$, a real-space grid with 301 grid-points and equidistant spacing $\Delta x = 0.1~a.u.$. For Fig.~2, the system is initialized in the ground state, perturbed by a linear response kick $v_{\text{kick}}(x,t)=- 10^{-6} x /\pi[(t-1)^2 + 10^{-4}]$ (in a.u.) and propagated for $4000$~a.u. of time. We extended the propagation time by a factor of 5 for the EIT curve ($g/\hbar\omega_c=0.01$) in order to improve the frequency resolution and clearly resolve the dip to 0 absorption.

\subsection{Perturbative to non-perturbative emission limit}

In order to numerically differentiate between the physical lifetime and the artificial width of excitations in real-time TDDFT due to finite propagation time, the damping of the dipole has to be considerable. For weakly confined electromagnetic environments, the lifetime of small isolated atomic and molecular systems is substantially longer than typical simulation times in TDDFT and it is therefore advisable to deduce the  linewidth by selecting a larger $A^{-1}$ than physically available and extrapolating towards the correct value \cite{bustamante2021dissipative} or to alternatively follow the Casida linear-response approach. As we have seen in the main text, some systems involving plasmonic coupling or strongly confined waveguides can strongly enhance the emission characteristics and shorten the lifetimes.

\begin{figure}
\includegraphics[width=1.0\columnwidth]{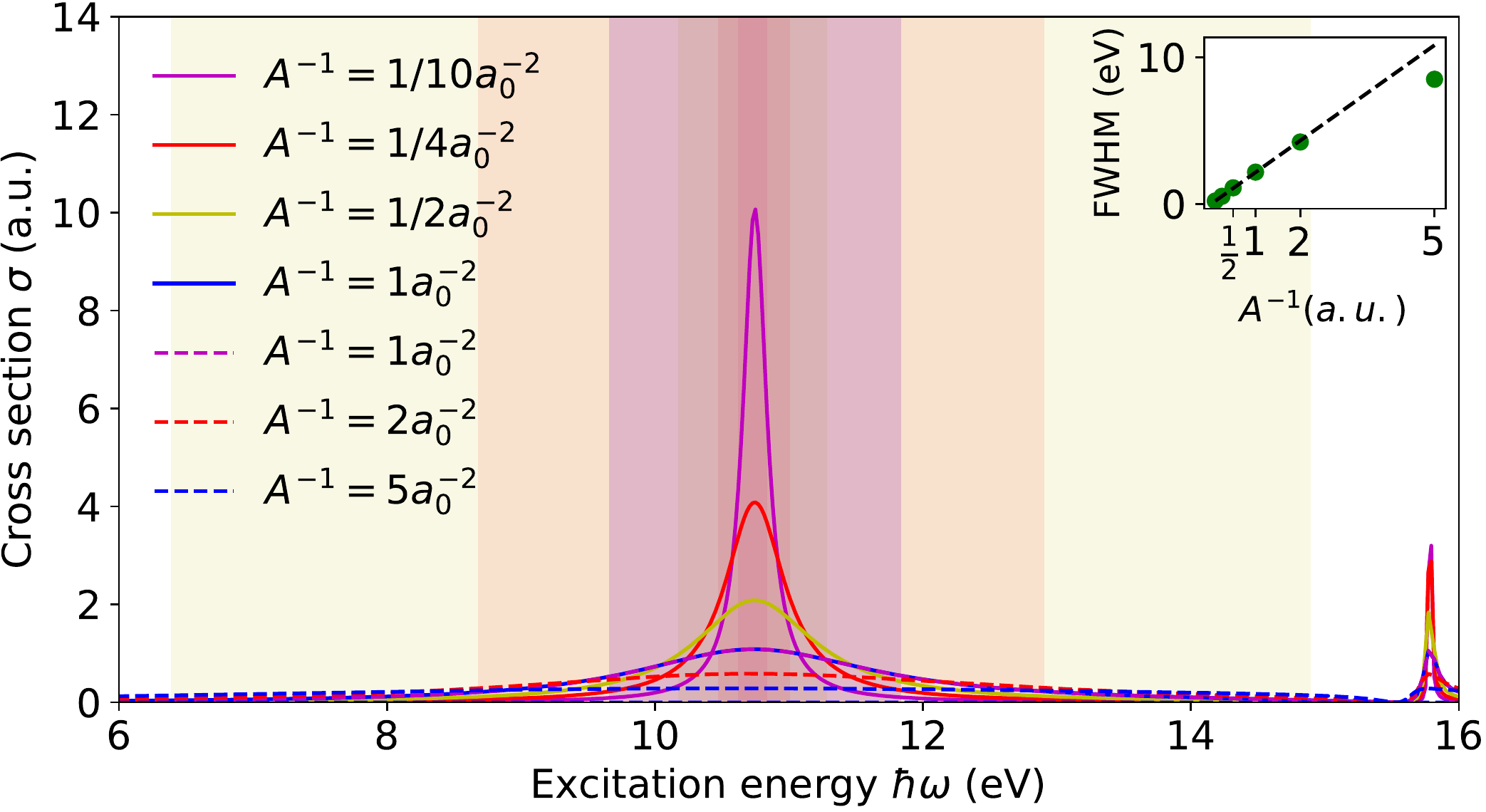}
\caption{Cross section for one-dimensional hydrogen with varying radiation-reaction strength. The FWHM (FWHM $ = 2\Gamma$) follows for a broad rage of emission values the perturbative treatment according to Wigner-Weisskopf illustrated as gray-dashed line.}
\label{fig:fwhm}
\end{figure}

Figure \ref{fig:fwhm} illustrates the cross-over between purely perturbative emission which follows Wigner-Weisskopf theory (gray dashed inset) and the non-perturbative limit in which the rate remains below the perturbative extrapolation. The good agreement in the perturbative limit is consistent with the computationally more involved explicit sampling QEDFT Casida description utilized by Flick et al. \cite{flick2018light}. As long as different excitations are clearly separated, the perturbative description provide an adequate estimate for the emission rate. Clearly hydrogen is hereby a pathological example due to the simplicity of the electronic structure and even when the linewidth deviates from the well defined Lorentzian, Wigner-Weisskopf remains a good estimate.

\subsection{Cross section}
The in fig.~2 provided photo-absorption cross-section $\sigma(\omega) = \frac{4\pi\omega}{c}\Im\alpha(\omega)$ (single dimension $\alpha(\omega) = \alpha_{xx}(\omega)$) has been calculated straight away from the definition $\alpha_{xx}(\omega) = R_x(\omega)/E_x(\omega)$. No additional window functions have been used in the FFT.

In addition to the 1D emission potential, we coupled a single cavity mode strongly to the hydrogen system. This mode was represented by $v_M(x t)=\sqrt{\frac{1}{\varepsilon_0 V}} e x \int_0^t dt' \cos(\omega (t-t')) \sqrt{\frac{1}{\varepsilon_0 V}} \dot{x}$ with $\hbar\omega=10.746~eV$ and V chosen according to the indicated $g/\hbar\omega$.

\subsection{Na chain}
\begin{figure}
\includegraphics[width=1.0\columnwidth]{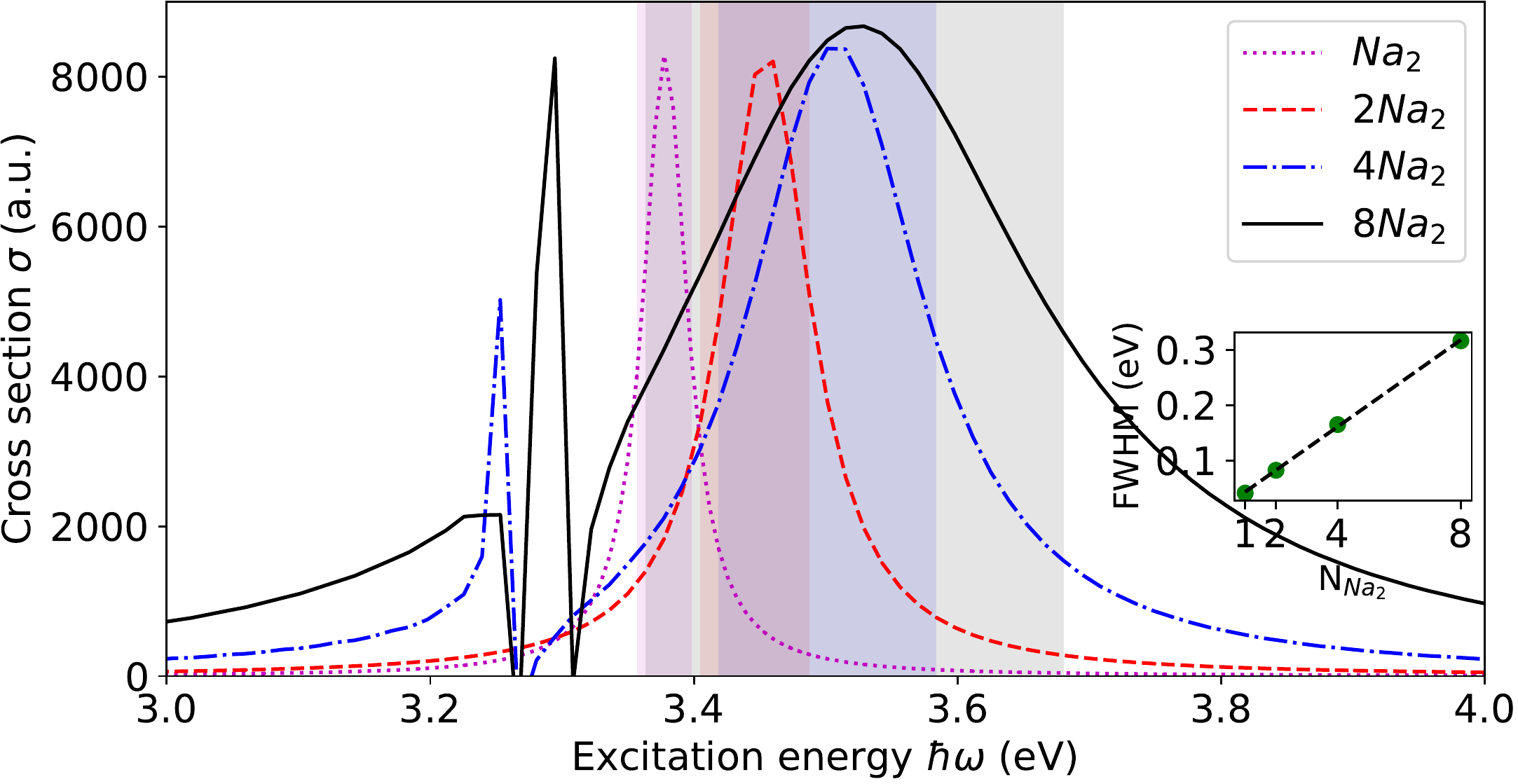}
\caption{Photoabsorption cross-section $\sigma_{zz}(\omega)$ for chains with variable length of far separated Na$_2$. The dimer axis is oriented along z with a bond distance of $1.104$~\AA, the chain along x has separations of $8$~\AA. We used the radiation-reaction potential as introduced previously with a quantization area of $35.05$~\AA$^2$ and polarization along z.}
\label{fig:superradiantSI}
\end{figure}

Fig.~4 has been obtained by perturbing the ground state by a linear response kick of strength $10^{-5}$ (ASE units).  The ground state and basis has been obtained in a simulation box of dimensions 96 \AA ~x 12 \AA ~x 12 \AA. The LDA functional, the standard Poisson solver, removing the momenta 1, 3 and 5, as well as a grid-spacing of 0.3 \AA~ and 12 (6 for $Na_2$) bands have been used.
We propagate (60000 for $Na_2$, 30000 for all others) steps with a time-step of 10 \AA$\sqrt{u/eV}$ and the double zeta polarized LCAO basis.
Fig.~\ref{fig:superradiantSI} shows a wider energy-window. Due to Coulomb-mediated dipole-dipole interactions in the chain, sharp low-energy peaks can be observed in addition to the blue-shift typical for H-aggregates. The overlap of low-energy and broadened main peaks results in Fano-like line-shapes.

\subsection{Al$_{201}$C$_{6}$H$_{6}$}
We use the structures and parameters published in \cite{rossi2019strong} (structure name Al201+b1,0) but extended the propagation time for benzene by a factor of 4 and the ones including the cluster by factor of 2. The benzene spectra have been smoothed via convolution with a sharp Lorentzian  of width $0.01$~eV while the spectra including the cluster used a width of $0.02$~eV to reduce remaining Sinc-contamination.
One should note here that without radiation-reaction, the broadening is the consequence of many discrete excitation peaks (delta-like for infinite propagation time) which provide in combination the impression of a broadened resonance.

\bibliography{sponemission} 